\documentclass[12 pt]{article}      
\usepackage{amsmath}
\usepackage{amsfonts}
\usepackage{graphicx}
\usepackage[table,xcdraw]{xcolor}
\usepackage{float}
\usepackage[myheadings]{fullpage}
\usepackage{fancyhdr}
\usepackage{lastpage}
\usepackage{graphicx, wrapfig, subcaption, setspace, booktabs}
\usepackage[T1]{fontenc}
\usepackage[font=small, labelfont=bf]{caption}
\usepackage{fourier}
\usepackage[protrusion=true, expansion=true]{microtype}
\usepackage[english]{babel}
\usepackage{sectsty}
\usepackage{url, lipsum}

\newcommand{\HRule}[1]{\rule{\linewidth}{#1}}
\begin{document}
\pagenumbering{gobble}
\clearpage
\begin{titlepage}
\title{ \normalsize \textsc{Master Sciences et Technologies \\ Specialit\'e "Probabilit\'es et Finance"}
        \\ [2.0cm]
        \textbf{Universit\'e Pierre et Marie Curie (Paris 6)}\\
        cohabilit\'e avec\\
        l' \textbf{\'Ecole Polytechnique}\\
        en collaboration avec\\
       l' \textbf{E.S.S.E.C.}\\
       l' \textbf{\'Ecole Normale Sup\'erieure (E.N.S. Ulm)}\\[2.0cm]
        \HRule{0.5pt} \\
        \LARGE \textbf{\uppercase{Memoire de stage de fin d'\'etudes}}
        \HRule{2pt} \\ [0.5cm]
        \normalsize  \vspace*{5\baselineskip}}

\date{ September 26, 2016}

\author{
        FIORIN Lucio \\
        Num\'ero \'etudiant: 3503139 \\ 
        }
        
\end{titlepage}
\maketitle
\newpage
\phantom{bla}
\newpage
\tableofcontents
\newpage
\pagenumbering{arabic}
\section*{Description of the company}
During my internship, I was part of the Quantitative team of the Soci\'et\'e G\'en\'erale Cross Asset Research division. I was based in London, while other parts of the team were in Paris, Hong Kong and Bangalore.\\
The internship period was from 2016, May $2^{nd}$ to 2016, November, $4^{th}$. My supervisors were Julien Turc, the head of the team, based in Paris, and Lorenzo Ravagli, based in London.\\
The division focuses on equity, cross asset, commodities, economic, credit analysis, fixed income and forex research. It employs a Cross-Asset Research approach involving analysis of the impact of major events on different asset classes, assessment of the links between asset classes, and strategic synthesis of this key information to provide an integrated view. The firm has global expertise on economics, quantitative research, strategy and asset allocation, commodities, credit, currencies, rates, emerging markets, equity, derivatives, technical analysis and Socially-Responsible Investment. The team offers daily, weekly and monthly multi asset research publications.\\
    It offers reports dedicated to the entire asset class range, as well as regional and thematic focused reports. The firm caters to asset managers. Soci\'et\'e G\'en\'erale Cross Asset Research is based in Paris, France with additional offices in Bangalore, India; Hong Kong, Hong Kong; London, The United Kingdom; Madrid, Spain; Milan, Italy; Moscow, Russia; New York, New York; Prague, Czech Republic; Singapore; Tokyo, Japan and Warsaw, Poland.
    \begin{figure}[H]
  \includegraphics[width=\linewidth]{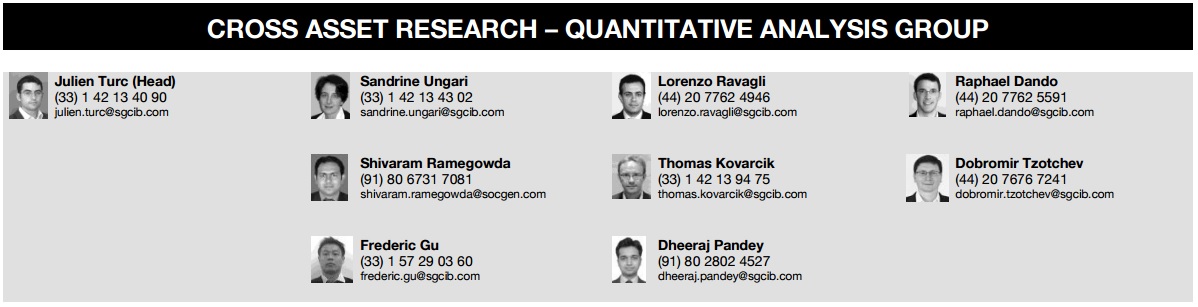}
\end{figure}
    \begin{figure}[H]
  \includegraphics[width=\linewidth]{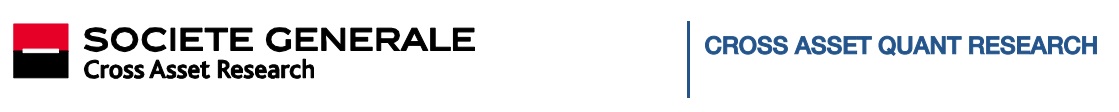}
\end{figure}
\newpage
\begin{figure}[H]
  \includegraphics[width=\linewidth]{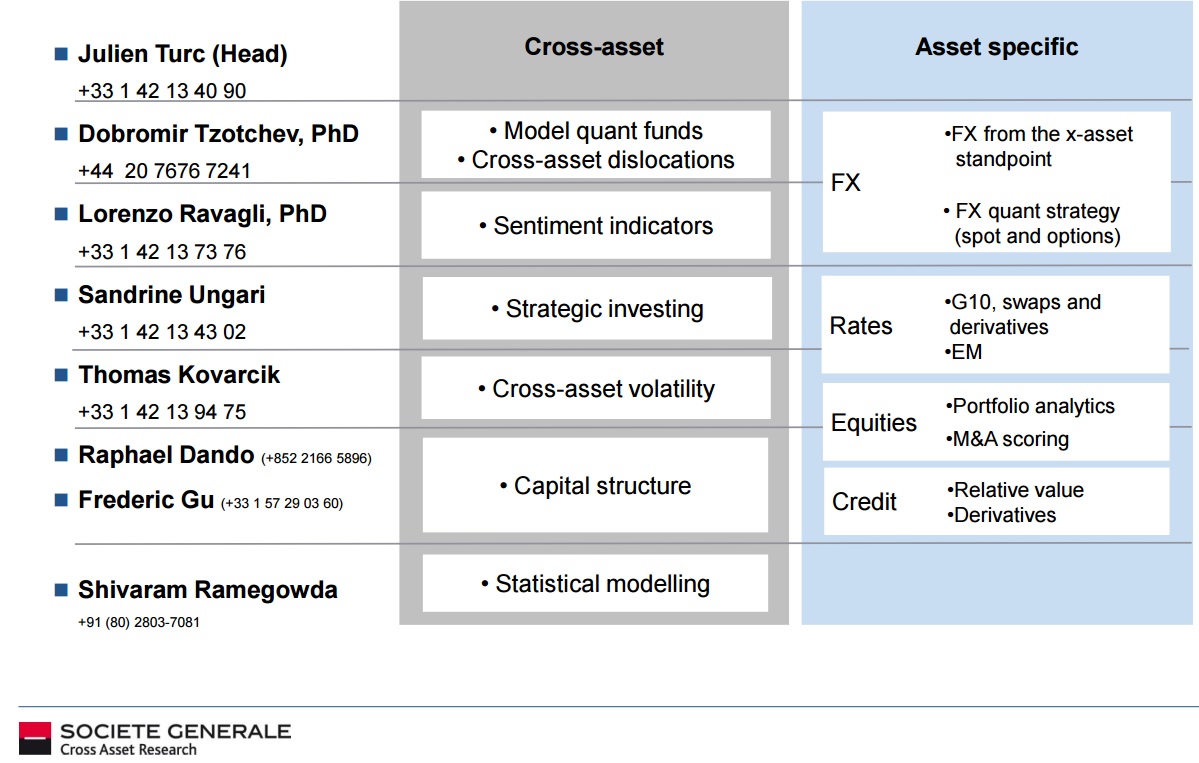}
\end{figure}
    
\phantom{bla}
\newpage

\section{Estimation of Historical volatility}
\subsection{Introduction}
A common comparison in the banking system is between the implied volatility of an option and the historical / realized volatility of the underlying. This is a useful tool that gives a hint on the fact that the market is pricing in a good way the option, with respect to the behavior of the underlying in the physic world. \\
Basically, if the implied volatility is higher than the historical volatility, it means that the derivative is more expensive then what it should be, so selling it would be a good trading choice, while if the implied volatility is lower than the historical volatility, the derivative is underestimated by the market, so buying it would be a good strategy. \\
This trading strategy, which works only on single assets, is not something completely formal or based on technical conditions, but is able to give an immediate and easy to understand trading opportunity, based only on what the market price is and what the behavior of an asset in the real world is.\\
While there is only one implied volatility, which is given by the market, there are many different techniques for the estimation of the historical volatility. This fact raises a question on which is the best method for the estimation of the volatility, and what properties should a good estimator have.\\
Recently, the development of the IT infrastructure in the banking world has given a lot of popularity and interest to the High Frequency environment. Nowadays, big prices database can be used in real time, and more information usually means a better estimation of the desired quantity.\\
Nevertheless, in this section we will not focus on the High Frequency estimation of historical volatility, but we will study different techniques which use intra day prices. We will compare them with more common techniques, as the numerous modifications of the autoregressive GARCH model, and we will focus on some of the properties of these estimators.\\
This comprehensive study could be also used in the view of forecasting of volatility, even if the possibility of predict volatility is out of the scope of this part, and so will not be developed.\\
The number of historical days for the historical volatility calculation changes the calculation, in addition to the estimate of the drift (or average amount stocks are assumed to rise). There should, however, be no difference between the average daily or weekly historical volatility. We also examine different methods of historical volatility calculation, including close-to-close volatility and exponentially weighted volatility, in addition to advanced volatility measures.\\
A comprehensive study of this topic can be found in \cite{Tsay} and \cite{Reider}
\subsection{First assumptions}
Volatility is defined as the annualized standard deviation of log returns, and the usual measure for historical volatility is given by the log returns from the close-to-close prices.\\
The calculation for standard deviation calculates the deviation from the average log return (or drift). This average log return has to be estimated from the sample, which can cause problems if the return over the period sampled is very high or negative. As over the long term very high or negative returns are not realistic, the calculation of volatility can be corrupted by using the sample log return as the expected future return.\\
In theory, the expected average value of an underlying at a future date should be the value of the forward at that date, because of non arbitrage conditions. As for all normal interest rates the forward return should be close to $100\%$ (for any reasonable sampling frequency i.e. daily/weekly/monthly). Hence for simplicity reasons it is easier to assume a zero log return as $\ln(100\%) = 0$. Since, in addition, returns are normally close to $100\%$ (the ratio between the value of today and the value of the previous trading day), the log returns are similar to the return minus $1$, due to the fact that $\ln(1+x) \sim x$ when $x$ is small.\\
If the return over the period is assumed to be the same for all periods, and if the mean return is assumed to be zero , that is confirmed by empirical data, since it is normally very close to zero, the standard deviation of the percentage change is simply the absolute value of the percentage return. Hence an underlying which moves $1\%$ has a volatility of $1\%$ for that period. As volatility is usually quoted on an annualized basis, this volatility has to be multiplied by the square root of the number of samples in a year, that is $252$ trading days in a year, the usual assumption in the industry.\\
While historical volatility can be measured monthly, quarterly or yearly it is usually measured daily or weekly. Normally, daily volatility is preferable to weekly volatility as $5$ times as many data points are available. However, if volatility over a long period of time is being examined between two different markets, weekly volatility could be the best measure to reduce the influence of different public holidays (and trading hours). If stock price returns are independent, then the daily and weekly historical volatility should on average be the same. If stock price returns are not independent, there could be a difference. Autocorrelation is the correlation between two different returns so independent returns have an autocorrelation of $0\%$.
\subsection{Volatility estimators}
\subsubsection{Moving Average}
The first and most simple estimator for the volatility that we will analyze is the moving average estimator (MA). This basic technique consists in using a moving window for the estimation of the volatility: at day $i$, the volatility is the standard deviation of the returns of the $N$ previous days. A common choice is $N=21$ days or $N=63$ days. The drawback of this estimator is the great instability in presence of shocks in the returns. Once a shock enters or leaves the moving window, there will be correspondingly a great volatility, even if in that precise day nothing has happened.\\
Because of this fact, and because of the lack of flexibility of this method, moving average techniques are not used in practice, but we will study them to see their behavior in different conditions.\\
When the window is as large as the data set, the moving average estimator becomes the constant volatility estimator.
\subsubsection{Autoregressive models}
The second model we will look at is the ARCH model, which stands for Auto Regressive Conditional Heteroscedasticity model. The AR comes from the fact that these models are autoregressive models in squared returns, which we will see later in this section. The conditional comes from the fact that, in these models, the information of the following period depends conditionally on the information present at this period. Heteroscedasticity means non constant volatility, which is a fundamental point.\\
Let us assume that the return on an asset is $r_t= \mu+ \sigma_t \varepsilon_t$, where $\mu$ is the mean return, $\sigma_t$ is the volatility at time $t$, and $\varepsilon_t$ is a sequence of standard Gaussian random variables, independent of $\sigma_t$. The residual return at time $t$ is $a_t  = r_t - \mu= \sigma_t \varepsilon_t$.\\
In the first ARCH model, developed in \cite{Engle1982}, the volatility is modeled in this way:
\begin{equation}
\sigma_t^2= \alpha_0+\alpha_1 a_{t-1}^2.
\end{equation}
We can see right away that a time varying $\sigma_t^2$ will lead to fatter tails, relative to a normal distribution, in the unconditional distribution of $a_t$.\\
In fact, it is easy to prove that the kurtosis of the residual returns is greater than $3$, that corresponds to the kurtosis of a Normal random variable:
$$
kurt(a_t )= \frac{\mathbb{E}[\sigma_t^4 ] \mathbb{E}[\varepsilon_t^4 ]}{(\mathbb{E}[\sigma_t^2 ] )^2 (\mathbb{E}[\varepsilon_t^2 ] )^2 }= 3 \frac{\mathbb{E}[\sigma_t^4 ]}{(\mathbb{E}[\sigma_t^2 ] )^2} >3,
$$
where the last inequality is given by the Jensen inequality, and
$$
Var(a_t )= \frac{\alpha_0}{1- \alpha_1 }
$$
In an ARCH(1) model, the variance of the next period only depends on the squared residual of the last period so a crisis that caused a large residual would not have the sort of persistence that we observe after actual crises. This has led to an extension of the ARCH model to a GARCH, or Generalized ARCH model, first developed in \cite{Bollerslev1986}, which is similar in spirit to an ARMA model. In a GARCH(1,1) model,
\begin{equation}
\sigma_t^2= \alpha_0+\alpha_1 a_{t-1}^2+\beta_1 \sigma_{t-1}^2.
\end{equation}
Since the unconditional variance of returns is
$$
\mathbb{E}(\sigma_t^2 )= \frac{\alpha_0}{1- \alpha_1-\beta_1 },
$$
we can write the GARCH(1,1) equation yet another way:
$$
\sigma_t^2=(1- \alpha_1-\beta_1 ) \mathbb{E}(\sigma^2)+ \alpha_1 a_{t-1}^2+\beta_1 \sigma_{t-1}^2.
$$
Written in this form, it is easy to see that next period's conditional variance is a weighted combination of the unconditional variance of returns, $\mathbb{E}(\sigma^2)$, squared residuals of the last period, $a_{t-1}^2$ , and the conditional variance of the last period, $\sigma_{t-1}^2$ , with weights $(1- \alpha_1-\beta_1 ), \alpha_1, \beta_1$ which sum to one.\\

An alternative measure could be to use an exponentially weighted moving average model, which is shown below. 
$$
\sigma_t^2= \lambda \sigma_{t-1}^2+ (1- \lambda) a_{t-1}^2.
$$
The parameter $\lambda$ is between $0$ (effectively $1$ day volatility) and $1$ (ignore current vol and keep vol constant). Normally, values of $0.94$ or $0.97$ are used.\\
Exponentially weighted volatilities are rarely used, partly due to the fact they do not handle regular volatility driving events such as earnings very well. Previous earnings jumps will have least weight just before an earnings date , when future volatility is most likely to be high, and most weight just after earnings, when future volatility is most likely to be low. It could, however, be of some use for indexes and equities.\\
Exponential weighted volatility has the advantage over standard historical volatility in that the effect of a spike in volatility gradually fades, as opposed to suddenly disappearing causing a collapse in historic volatility. For example, if we are looking at the historical volatility over the past month and a spike in realized volatility suddenly occurs the historical volatility will be high for a month, then collapse. Exponentially weighted volatility will rise at the same time as historical volatility, and then gradually decline to lower levels, arguably in a similar way to how implied volatility spikes, then mean reverts.\\
\subsubsection{Modification of the GARCH model}
One empirical observation is that, in many markets, the impact of negative price moves on future volatility is different from that of positive price moves. This is particularly true in equity markets.\\
It is clear that not only does the magnitude of $a_t^2$ affects future volatility, but the sign of the residual return also affects future volatility, at least for equities. It is not clear why volatility should increase more when the level of stock prices drop compared to a stock price rise. In fact, on one hand, as stocks drop, the debt/equity ratios increase and stocks become more volatile with higher leverage ratios. But the changes in volatility associated with stock market drops are much larger than that which could be explained by leverage alone. One model which can take in account this asymmetry is the Threshold GARCH (TGARCH) model, also known as the GJR-GARCH model, based on the work in \cite{Runkle1993}:
\begin{equation}
\sigma_t^2= \alpha_0+\alpha_1 a_{t-1}^2+\gamma_1 1_{a_{t-1}^2<0} a_{t-1}^2+\beta_1 \sigma_{t-1}^2.
\end{equation}
This method is of course more flexible than the original GARCH model, but it is also more difficult to estimate, since we add a parameter in the model.\\
Another variant on GARCH to account for the asymmetry between up and down moves described above is the EGARCH model of \cite{Nelson1991}. In an EGARCH
\begin{equation}
\ln(\sigma_t^2)=\alpha_0+\frac{\alpha_1 a_{t-1}^2+\gamma_1 |a_{t-1}  |}{\sigma_{t-1}}+\beta_1 \ln(\sigma_{t-1}^2).
\end{equation}
There are different other variations of asymmetric GARCH models which can be found in the literature, but we will focus only on these ones, analyzing their properties.\\
\subsubsection{Advanced volatility measures}
The increasing availability of high-frequency price time series has permitted the development of new integrated variance estimators that are more efficient than the realized variance estimator.\\
Close-to-close volatility is usually used as it has the benefit of using the closing auction prices only. Should other prices be used, then they could be vulnerable to manipulation or a "fat fingered" trade. However, a large number of samples need to be used to get a good estimate of historical volatility, and using a large number of closing values can obscure short-term changes in volatility. There are, however, different methods of calculating volatility using some or all of the open (O), high (H), low (L) and close (C).\\
For most markets, intra day volatility is greatest just after the open (as results are often announced around the open) and just before the close (performance is often based upon closing prices). Intra day volatility tends to sag in the middle of the day, due to the combination of a lack of announcements and reduced volumes/liquidity due to lunch breaks. For this reason using an estimate of volatility more frequent than daily tends to be very noisy. Traders who wish to take into account intra day prices should instead use an advanced volatility measure.\\
Before introducing new volatility measures, it is important also to have a measure of the goodness of the estimation. The efficiency is the number used to test this property, and it is defined as the variance of the most basic estimator, the close to close,  divided by the variance of the new estimator:
$$
eff_{\text{estimator}} = \frac{ var\left ( \sigma_{\text{close to close}} \right)}{ var \left( \sigma_{\text{estimator}} \right) }. 
$$ 
Since the aim of the estimators is to have the biggest efficiency, we are looking for low variance estimators.\\
A comprehensive study of this estimators can be found in \cite{Tsay}, \cite{Saichev} and \cite{Bernett}.
In the following, $c_i,l_i,h_i,o_i$ are, respectively, the closing, low, high and closing price at day $i$, $N$ is the number of trading days considered.\\
\begin{itemize}
	\item \textbf{Close to close (C)}: The most common type of calculation that benefits from only using reliable prices from closing auctions. By definition its efficiency is $1$ at all times. 
\begin{equation}
\sigma_{\text{close to close}} = \sqrt{\frac{1}{N}} \sqrt{ \sum_{i=1}^{N} \left( \ln \left( \frac{c_i}{c_{i-1}} \right)^2 \right) }.
\end{equation}
\item \textbf{Parkinson (HL), see \cite{Parkinson}}: The first advanced volatility estimator was created by Parkinson in 1980, and instead of using closing prices it uses the high and low price. One drawback of this estimator is that it assumes continuous trading, hence it underestimates the volatility as potential movements when the market is shut are ignored. As this estimate only uses the high and low price for an underlying, it is less sensitive to differences in trading hours. For example, as the time of the EU and US closes are approximately half a trading day apart, they can give very different returns. Using the high and low means the trading over the whole day is examined, and the days overlap. As it does not handle jumps, on average it underestimates the volatility, as it does not take into account highs and lows when trading does not occur (weekends, between close and open). Although it does not handle drift, this is usually small. The Parkinson estimate is up to $5.2$ times more efficient than the close to close estimate. While other measures are more efficient based on simulated data, some studies have shown it to be the best measure for actual empirical data.
\begin{equation}
\sigma_{\text{Parkinson}} = \sqrt{\frac{1}{N}} \sqrt{ \frac{1}{4 \ln 2 } \sum_{i=1}^{N} \left( \ln \left( \frac{h_i}{l_{i}} \right) \right)^2 }.
\end{equation}
\item \textbf{Garman-Klass (OHLC), see \cite{GarmanKlass}}: Later in 1980 the Garman-Klass volatility estimator was created. It is an extension of Parkinson which includes opening and closing prices. If opening prices are not available the close from the previous day can be used instead. As overnight jumps are ignored the measure underestimates the volatility. This estimate is the most powerful for stocks with Brownian motion, zero drift and no opening jumps, i.e. opening price is equal to closing price of previous period. While it is up to $7.4$ times as efficient as the close to close estimate, it also underestimates the volatility, as like Parkinson it assumes no jumps. 
\begin{equation}
\sigma_{\text{Garman-Klass}} = \sqrt{\frac{1}{N}} \sqrt{ \frac{1}{ 2 } \sum_{i=1}^{N} \left( \ln \left( \frac{h_i}{l_{i}} \right) \right)^2  - (2 \ln 2) \left( \ln \left(\frac{c_i}{o_i} \right) \right)^2 }.
\end{equation}
\item \textbf{Rogers-Satchell (OHLC), see \cite{RogersSatchell}}: Securities which have a drift, or non-zero mean, require a more sophisticated measure of volatility. The Rogers-Satchell volatility created in the early 1990s is able to properly measure the volatility for securities with non-zero mean. It does not, however, handle jumps, hence it underestimates the volatility. The efficiency of the Rogers-Satchell estimate is similar to that for Garman-Klass, however it benefits from being able to handle non-zero drift. Opening jumps are, however, not handled well.
\begin{equation}
\sigma_{\text{Rogers-Satchell}} = \sqrt{\frac{1}{N}} \sqrt{ \sum_{i=1}^{N}  \ln \left( \frac{h_i}{c_{i}} \right) \ln \left( \frac{h_i}{o_{i}} \right) + \ln \left( \frac{l_i}{c_{i}} \right) \ln \left( \frac{l_i}{o_{i}} \right)  }.
\end{equation}
\item \textbf{Garman-Klass Yang-Zhang extension (OHLC), see \cite{YangZhang}}: Yang-Zhang modified the Garman-Klass volatility measure in order to let it handle jumps. The measure does assume a zero drift, hence it will overestimate the volatility if a security has a non-zero mean return. As the effect of drift is small, the fact continuous prices are not available usually means it underestimates the volatility, but by a smaller amount than the previous alternative measures. It has an efficiency of $8$ times the close to close estimate. 
\begin{equation}
\sigma_{\text{GKYZ}} = \sqrt{\frac{1}{N}} \sqrt{ \sum_{i=1}^{N} \left( \ln \left( \frac{o_i}{c_{i-1}} \right) \right)^2  + \frac{1}{2} \left( \ln \left(\frac{h_i}{l_i} \right) \right)^2 - (2 \ln 2) \left( \ln \left(\frac{c_i}{o_i} \right) \right)^2 }.
\end{equation}
\item \textbf{Yang-Zhang (OHLC), see \cite{YangZhang}}: The most powerful volatility estimator which has minimum estimation error. In 2000 Yang-Zhang created a volatility measure that handles both opening jumps and drift. It is the sum of the overnight volatility (close-to-open volatility) and a weighted average of the Rogers-Satchell volatility and the open-to-close volatility. The assumption of continuous prices does mean the measure tends to slightly underestimate the volatility.  It is up to a maximum of 14 times as efficient as the close to close estimate.
\begin{equation}
\sigma_{\text{Yang-Zhang}} = \sqrt{ \sigma^2_{\text{overnight}} + k \sigma^2_{\text{open to close}} + (1 - k) \sigma^2_{\text{Rogers - Satchell} }},
\end{equation}
where
$$
\sigma^2_{\text{overnight}} = \frac{1}{N-1} \sum_{i=1}^N \left( \ln \left( \frac{o_i}{c_{i-1}} \right) - \overline{\ln \left( \frac{o_i}{c_{i-1}} \right)} \right)^2,
$$
$$
\sigma^2_{\text{open to close}} = \frac{1}{N-1} \sum_{i=1}^N \left( \ln \left( \frac{c_i}{o_{i}} \right) - \overline{\ln \left( \frac{c_i}{o_{i}} \right)} \right)^2,
$$
$$
k = \frac{10.34}{1.34 + \frac{N+1}{N-1}}.
$$
\end{itemize}
While efficiency (how volatile the measure is) is important, so too is the bias of the estimator. Bias depends on the sample size, and the type of distribution the underlying security has.\\
Generally, the close-to-close volatility estimator is too big, since it does not model overnight jumps, while alternative estimators are too small, as they assume continuous trading, and discrete trading will have a smaller difference between the maximum and minimum.\\
The key variables which determine the bias are:
\begin{itemize}
\item Sample size: As the standard close-to-close volatility measure suffers with small sample sizes, this is where alternative measures perform best. 
\item	Volatility of volatility: While the close-to-close volatility estimate is relatively insensitive to a changing volatility (vol of vol), the alternative estimates are far more sensitive. This bias increases the more vol of vol increases , i.e. more vol of vol means a greater underestimate of volatility.
\item	Overnight jumps between close and open: Approximately one-sixth of equity volatility occurs outside the trading day. Overnight jumps cause the standard close-to-close estimate to overestimate the volatility, as jumps are not modeled. Alternative estimates which do not model jumps (Parkinson, Garman Klass and Rogers-Satchell) underestimate the volatility. Yang-Zhang estimates, both YangZhang extension of Garman Klass and the Yang-Zhang measure itself, will converge with standard close-to-close volatility if the jumps are large compared to the overnight volatility. 
\item	Drift of underlying: If the drift of the underlying is ignored as it is for Parkinson and Garman Klass, and the Yang Zhang extension of Garman Glass, then the measure will overestimate the volatility. This effect is small for any reasonable drifts, i.e. if we are looking at daily, weekly or monthly data. 
\item	Correlation daily volatility and overnight volatility: While Yang-Zhang measures deal with overnight volatility, there is the assumption that overnight volatility and daily volatility are uncorrelated. Yang-Zhang measures will underestimate volatility when there is a correlation between daily return and overnight return, and vice versa, but this effect is small.
\end{itemize}
\subsection{A case study}
While all the works in literature focus on the statistical, and so theoretical, efficiency of the various estimators, we study the autocorrelation of the residuals in order to test which method gives a better non correlated volatility series. All the figures and the tables here are related on the SPX. The Standard \& Poor's 500, often abbreviated as the S\&P 500, or just "the S \& P", is an American stock market index based on the market capitalizations of 500 large companies having common stock listed on the NYSE or NASDAQ. Data are from January 1999 to June 2016, and are downloaded from Bloomberg.\\
We introduce different types of test, which are based on the estimated residuals
$$
res_{t,i} = \frac{ret_{t}}{\sigma_{t,i}},
$$
where $i$ is the method chosen, $ret_{t} $ is the series of the log returns of the spot price at time $t$, and $\sigma_{t,i}$ is the value of the volatility estimator at time $t$:
\begin{enumerate}
\item	Constant volatility
\item	Parkinson
\item   Garman Klass
\item	Rogers Satchell
\item	German Klass Yang Zhang
\item	Yang Zhang
\item	Moving Average 21 days
\item	Moving Average 63 days
\item	Garch (1,1)
\item	Asymmetric Garch (1,1)
\item	Exponential Garch (1,1)
\item	Exponential Weighted Moving Average (1,1)
\end{enumerate}
\subsubsection{Normality test} 
We first check the normality of the series of the residuals, studying the QQ-plot and the skewness and kurtosis parameters. Remember that, in the case of a Gaussian series, the skewness would be $0$ and the kurtosis equal to $3$. 
\begin{figure}[H]
  \includegraphics[width=\linewidth]{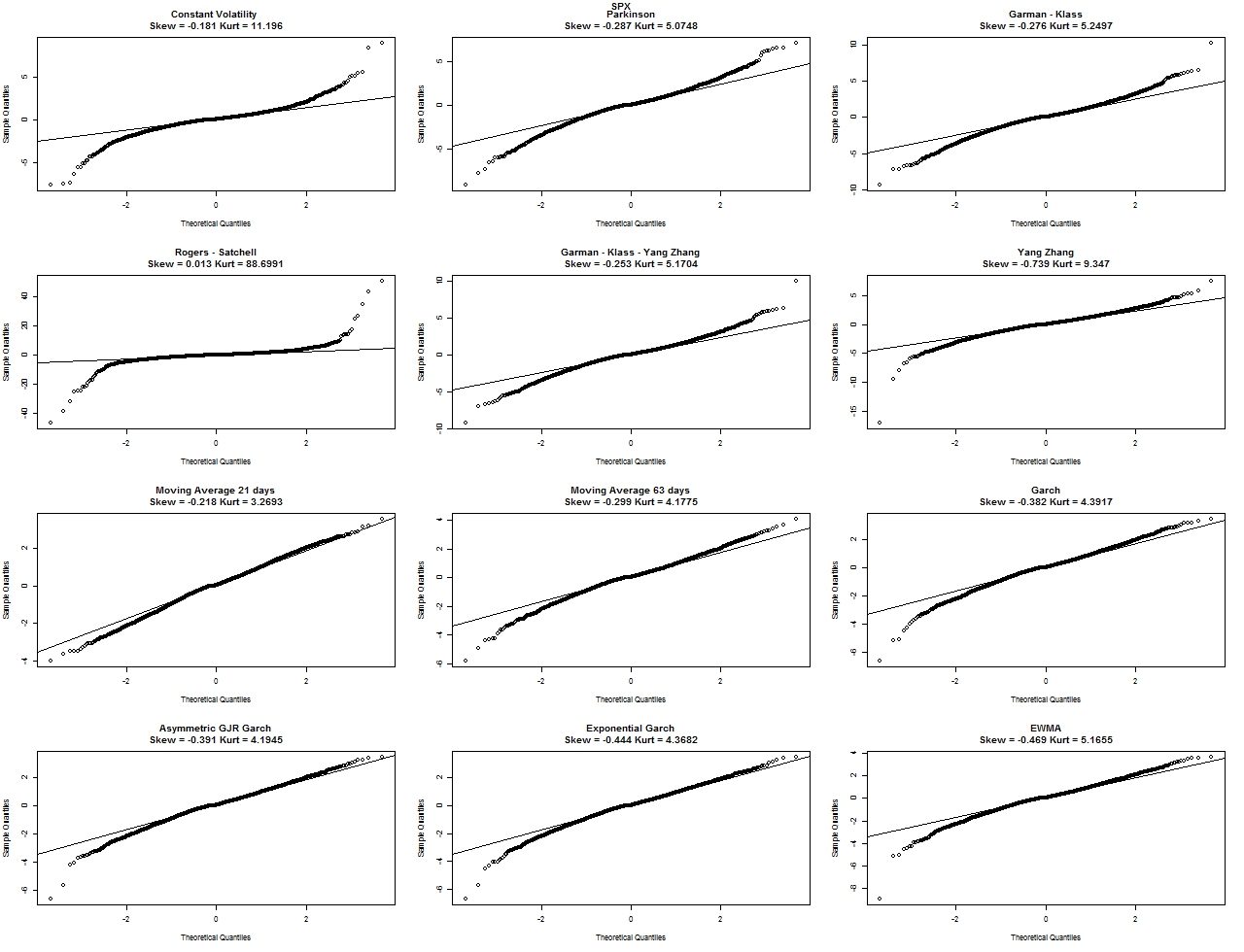}
  \caption{QQ plot, Skewness and Kurtosis parameters in the SPX asset for all the estimators.}
  \label{qqplot_SPX}
\end{figure}
We can immediately see that all the Garch estimators and the Moving Average estimators perform better in term of skewness and kurtosis. There are estimators that perform better in term of skewness (see the Rogers Satchell estimator), but really badly from a kurtosis point of view.\\
It is important to understand that the data frame is of 17 years. It could be an interval too important, but the interest of the study is to give a more general overview on the possible estimation techniques, rather than building or choosing the best estimator for a given asset in a given time period.\\
\subsubsection{Autocorrelation test}
We study then the behavior of the series of the residuals, testing the autocorrelation function. The autocorrelation of a random process is the correlation between values of the process at different times, as a function of the time lag. Since all the computations are done in \textit{R}, the default time lag is $10 \log_{10}N$, where $N$ is the number of observations. Assuming the hypothesis of the residuals to be close to white noise, we study the autocorrelation of the series of the residuals and of the squared of the residuals.
\begin{figure}[H]
  \includegraphics[width=\linewidth]{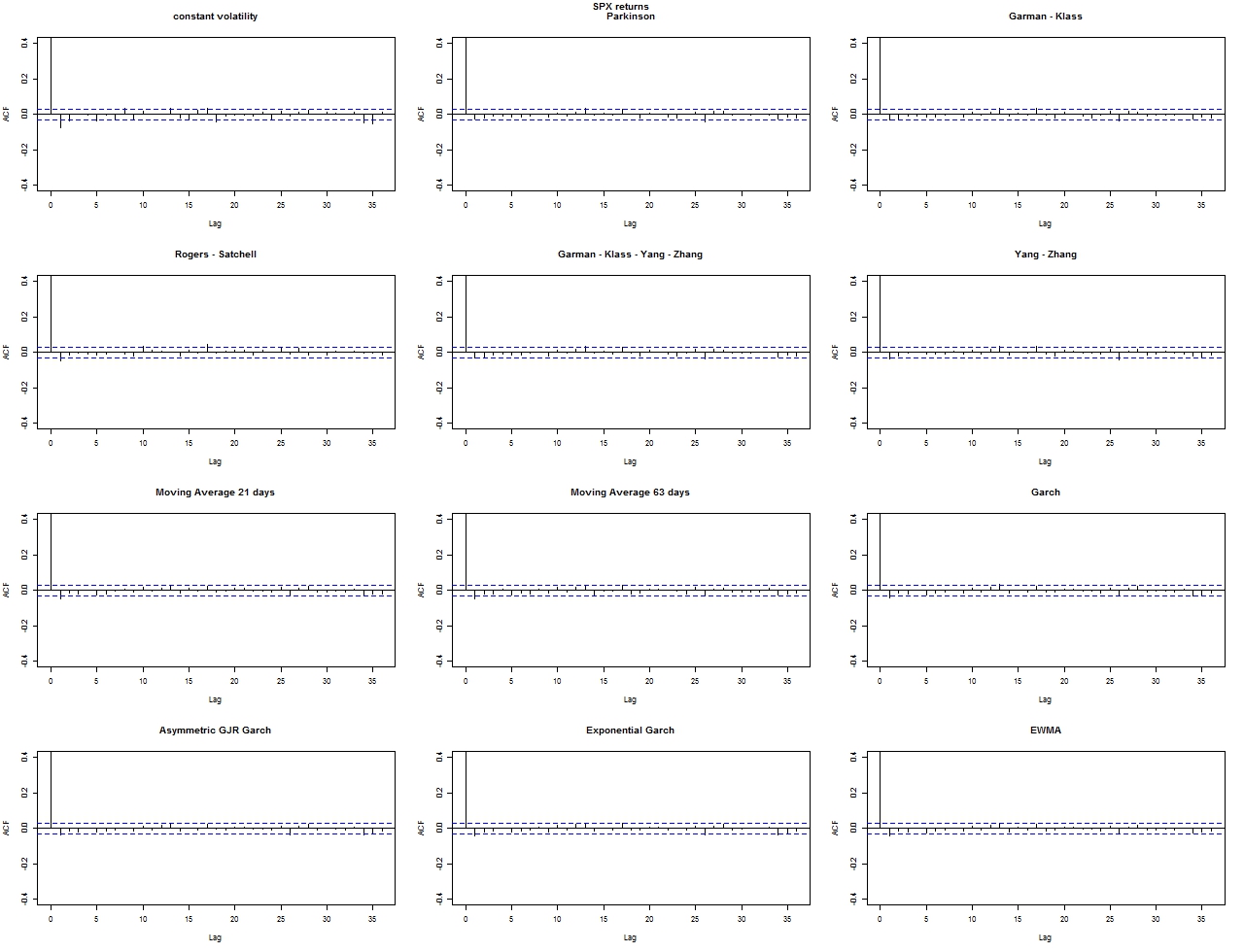}
  \caption{Autocorrelation test for the series of $res_{t,i}$ in the SPX asset for all the estimators.}
  \label{comparison_acf_SPX_returns}
\end{figure}
From the study of the autocorrelation of the returns in Figure \ref{comparison_acf_SPX_returns} we cannot have much information. The results are almost the same for every estimator, so this test does not give much information about the goodness of the methods.
\begin{figure}[H]
  \includegraphics[width=\linewidth]{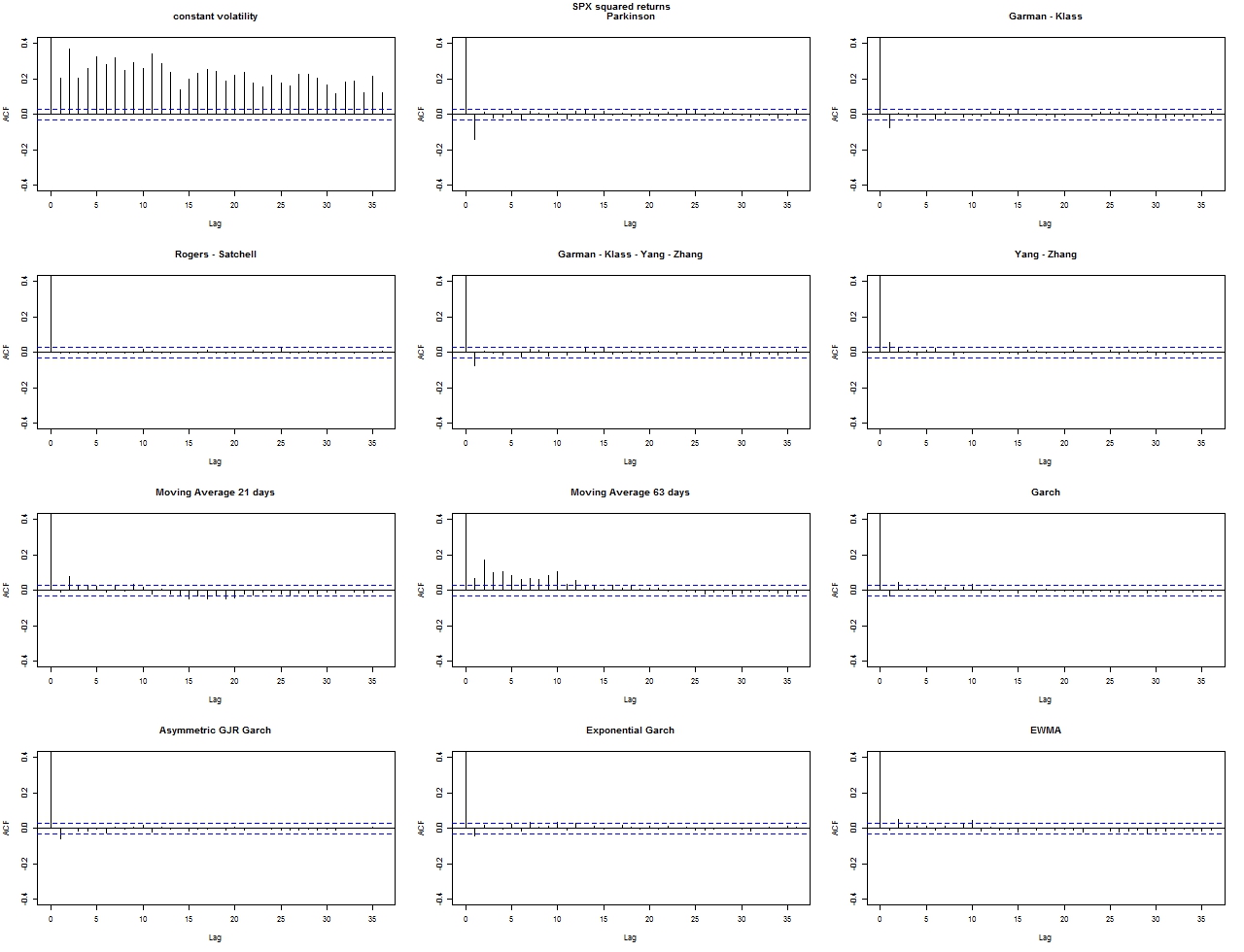}
  \caption{Autocorrelation test for the series of $(res_{t,i})^2$ in the SPX asset for all the estimators.}
  \label{comparison_acf_SPX_squaredreturns}
\end{figure}
If we focus on the autocorrelation of the squared returns in Figure \ref{comparison_acf_SPX_squaredreturns}, we see immediately how the constant volatility is, as highly expectable, not a good choice. Also the (static) moving average choice is really poor, since it appears that the autocorrelation is influenced by the natural construction of the estimator. Both advanced measures and GARCH based estimators seem promising from this aspect.\\
\subsubsection{Ljung - Box test}
The Ljung - Box test (named for Greta M. Ljung and George E. P. Box) is a type of statistical test of whether any of a group of autocorrelations of a time series are different from zero. Instead of testing randomness at each distinct lag, it tests the "overall" randomness based on a number of lags. \\
The Ljung - Box test may be defined as:\\
$H_0$: The data are independently distributed, i.e., the correlations in the population from which the sample is taken are $0$, so that any observed correlations in the data result from randomness of the sampling process.\\
$H_a$: The data are not independently distributed; they exhibit serial correlation.\\
The test statistic is:
$$
Q = N(N+2) \sum_{i=1}^h \frac{\hat{\rho}^2_i}{N - i},
$$
where $N$ is the sample size, $\hat{\rho}^2_i$ is the sample autocorrelation at lag $i$, and $h$ is the number of lags being tested. Under $H_0$ the statistic $Q$ follows a $\chi_{(h)}^2$ distribution. For significance level $\alpha$, the critical region for rejection of the hypothesis of randomness is
$$
Q > \chi^2_{1-\alpha, h},
$$
where $\chi^2_{1-\alpha, h}$ is the $\alpha$-quantile of the chi-squared distribution with $h$ degrees of freedom.\\
If the p value is under the threshold of $5\%$, the test accepts the hypothesis of autocorrelation of the time series. The test is performed with three different lags.
\begin{table}[H]
\centering

\begin{tabular}{l|ll|ll|ll}
         & lag = 10 & p value                      & lag = 15 & p value                      & lag = 20 & p value                      \\
         \hline
Constvol & 3505.63             & \cellcolor[HTML]{FD6864}0.00 & 4913.57             & \cellcolor[HTML]{FD6864}0.00 & 6086.93             & \cellcolor[HTML]{FD6864}0.00 \\
P        & 103.92              & \cellcolor[HTML]{FD6864}0.00 & 113.66              & \cellcolor[HTML]{FD6864}0.00 & 115.56              & \cellcolor[HTML]{FD6864}0.00 \\
GK       & 31.32               & \cellcolor[HTML]{FD6864}0.00 & 35.96               & \cellcolor[HTML]{FD6864}0.00 & 37.12               & \cellcolor[HTML]{FD6864}0.01 \\
RS       & 2.05                & \cellcolor[HTML]{67FD9A}1.00 & 2.42                & \cellcolor[HTML]{67FD9A}1.00 & 3.53                & \cellcolor[HTML]{67FD9A}1.00 \\
GKYZ     & 32.26               & \cellcolor[HTML]{FD6864}0.00 & 38.20               & \cellcolor[HTML]{FD6864}0.00 & 39.21               & \cellcolor[HTML]{FD6864}0.01 \\
YZ       & 21.71               & \cellcolor[HTML]{FD6864}0.02 & 22.38               & \cellcolor[HTML]{67FD9A}0.10 & 23.27               & \cellcolor[HTML]{67FD9A}0.28 \\
MA21days & 44.66               & \cellcolor[HTML]{FD6864}0.00 & 60.92               & \cellcolor[HTML]{FD6864}0.00 & 97.31               & \cellcolor[HTML]{FD6864}0.00 \\
MA63days & 415.37              & \cellcolor[HTML]{FD6864}0.00 & 438.82              & \cellcolor[HTML]{FD6864}0.00 & 446.16              & \cellcolor[HTML]{FD6864}0.00 \\
Garch    & 21.50               & \cellcolor[HTML]{FD6864}0.02 & 24.04               & \cellcolor[HTML]{67FD9A}0.06 & 24.64               & \cellcolor[HTML]{67FD9A}0.22 \\
GJRGarch & 22.43               & \cellcolor[HTML]{FD6864}0.01 & 25.78               & \cellcolor[HTML]{FD6864}0.04 & 26.72               & \cellcolor[HTML]{67FD9A}0.14 \\
eGarch   & 25.97               & \cellcolor[HTML]{FD6864}0.00 & 30.74               & \cellcolor[HTML]{FD6864}0.01 & 33.39               & \cellcolor[HTML]{FD6864}0.03 \\
EWMA     & 26.54               & \cellcolor[HTML]{FD6864}0.00 & 30.04               & \cellcolor[HTML]{FD6864}0.01 & 31.29               & \cellcolor[HTML]{67FD9A}0.05
\end{tabular}
\caption{Ljung - Box test with three different lags in the SPX asset for all the estimators. }\label{Table:LjungBox}
\end{table}
Analyzing Table \ref{Table:LjungBox}, a high $p$ value seems to indicate absence of correlation in the series of the residuals, but the test does not prove this fact. \\
We can see how the Moving Average estimation is really poor when it comes to analyzing the autocorrelation of the residuals. All the Advanced volatility measures, using intra day prices, perform quite well, and their ability to get uncorrelated residuals is very interesting.\\
Rogers Satchell seems to perform incredibly well, and also Yang Zhang is a good estimator, following the results of this test.
\subsubsection{Volatility of the estimators}
Even if the advanced estimators are constructed in order to reduce their own volatility with respect to the close to close variance, the fact that they not depend on the information of the past, but only on the prices of the day, make them more volatile than the autoregressive estimators, which are constructed in order to depend from the past in different ways.
\begin{figure}[H]
  \includegraphics[width=\linewidth]{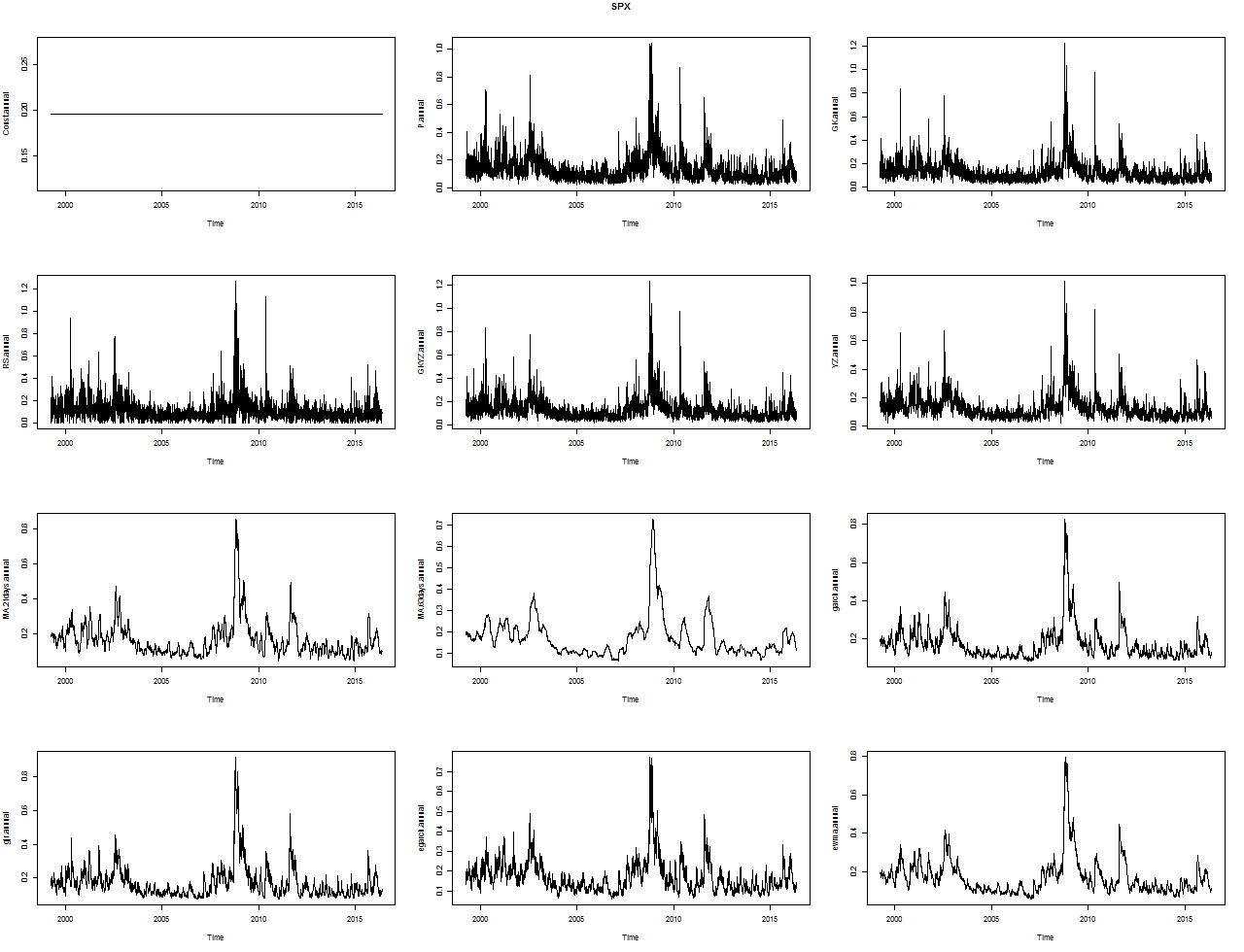}
  \caption{Time series of $\sigma_{t,i}$ in the SPX asset for all the estimators.}
  \label{series_SPX}
\end{figure}
We can immediately see from Figure \ref{series_SPX} that the advanced estimators are incredibly volatile, if compared with the Moving Average estimators and the GARCH estimators, since the last ones present a smoothing effect given by the regressive part. Inspired by the figure, we introduce a measure of volatility of the time series. Assuming Gaussian properties of the volatility series, we compute the standard deviation of the "log returns of the volatilities" for all estimators and all assets.\\

\begin{table}[H]
\centering

\begin{tabular}{c|c|c|c|c|c}
ConstVol & Parkinson  & Garman Klass & Rogers Satchell & GKYZ   & Yang Zhang \\
0.00                & 0.53       & 0.50         & 0.77            & 0.51   & 0.32       \\
\hline
MA 21 days         & MA 63 days & GARCH        & GJR GARCH       & eGARCH & EWMA       \\
0.06                & 0.02       & 0.06         & 0.08            & 0.09   & 0.05      
\end{tabular}
\caption{Standard deviation of the time series of $\sigma_{t,i}$ in the SPX asset for all the estimators.}\label{Table:volvol}
\end{table}

One drawback of Intra day estimators, that can be immediately seen from Table \ref{Table:volvol}, is that, even if the correlation of the residuals is very low, the volatility of the time series is very high, providing some issues in the eventual forecasting of the volatility series, even if this is not a matter in this study.
\subsection{GARCH and EWMA on a daily estimation of the parameters}
Going back to Autoregressive models, we analyze how the parameters of these autoregressive models are changing in time. Every calibration is done in an increasing time window, and the last fit contains all the data set. \\ 
Basically, the volatility at time $t$ is given by the last value of the calibrated series from day $1$ to day $t$. With this technique, every element of the time series of volatility is obtained from a different calibration.
\begin{figure}[H]
  \includegraphics[width=\linewidth]{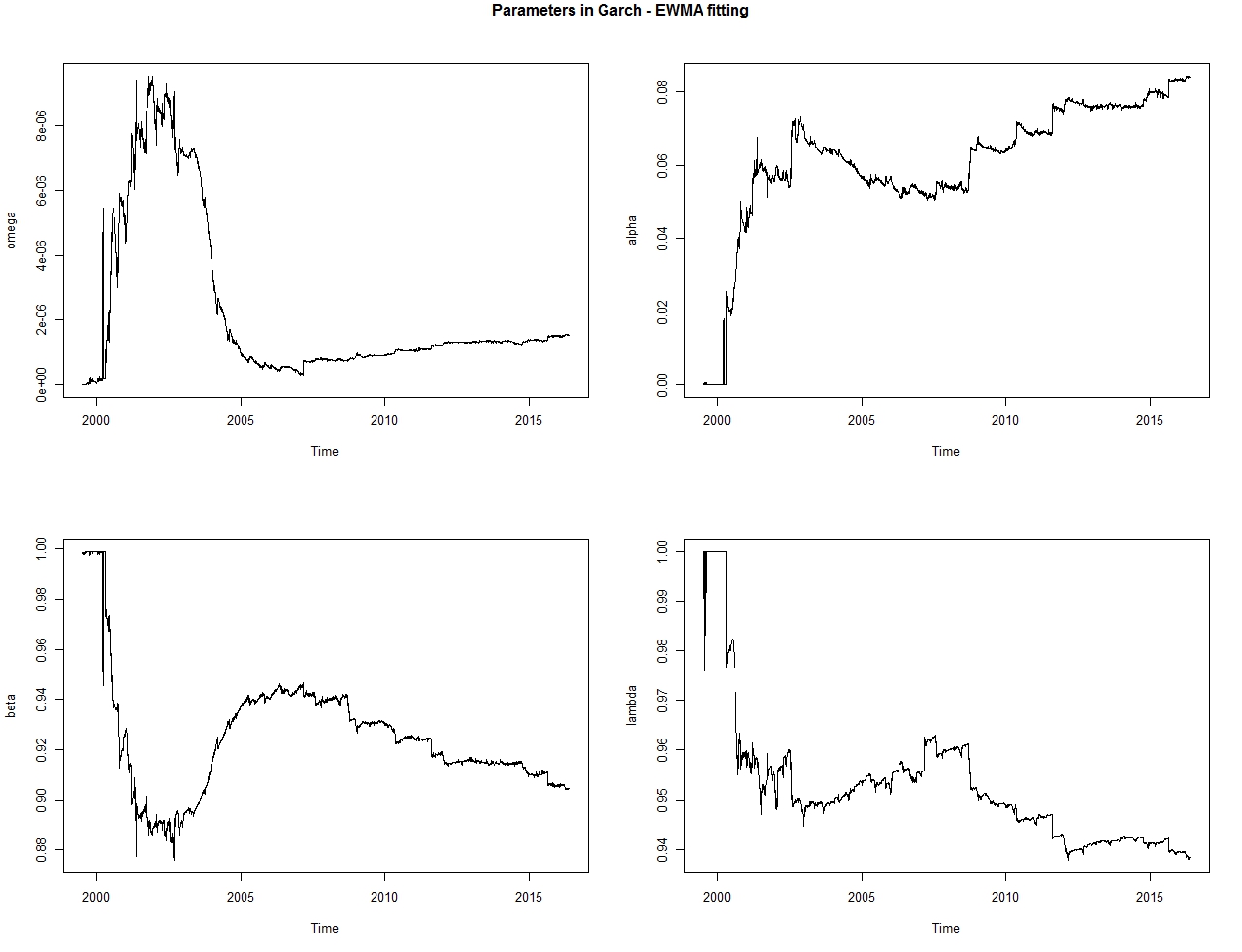}
  \caption{Time series of the parameters in a Garch and EWMA estimation of the volatility.}
  \label{garch-ewma}
\end{figure}
Figure \ref{garch-ewma} gives an idea of the importance to recalibrate the model to the data, and not to do it just once on the whole set. But it is also important to see that a daily recalibration is almost useless, since the parameters are not changing on a daily basis, but more on a monthly basis.\\
In Figure \ref{garch-ewma},  $\omega, \alpha$ and $\beta$ are the parameters of the GARCH model, while $\lambda$ is for the EWMA model.\\
If we test this new volatility series, where we use at every time step $t$ the final value of the fitting on the data from the beginning to $t$, we can see that the results on the squared residuals are not very encouraging
\begin{figure}[H]
  \includegraphics[width=\linewidth]{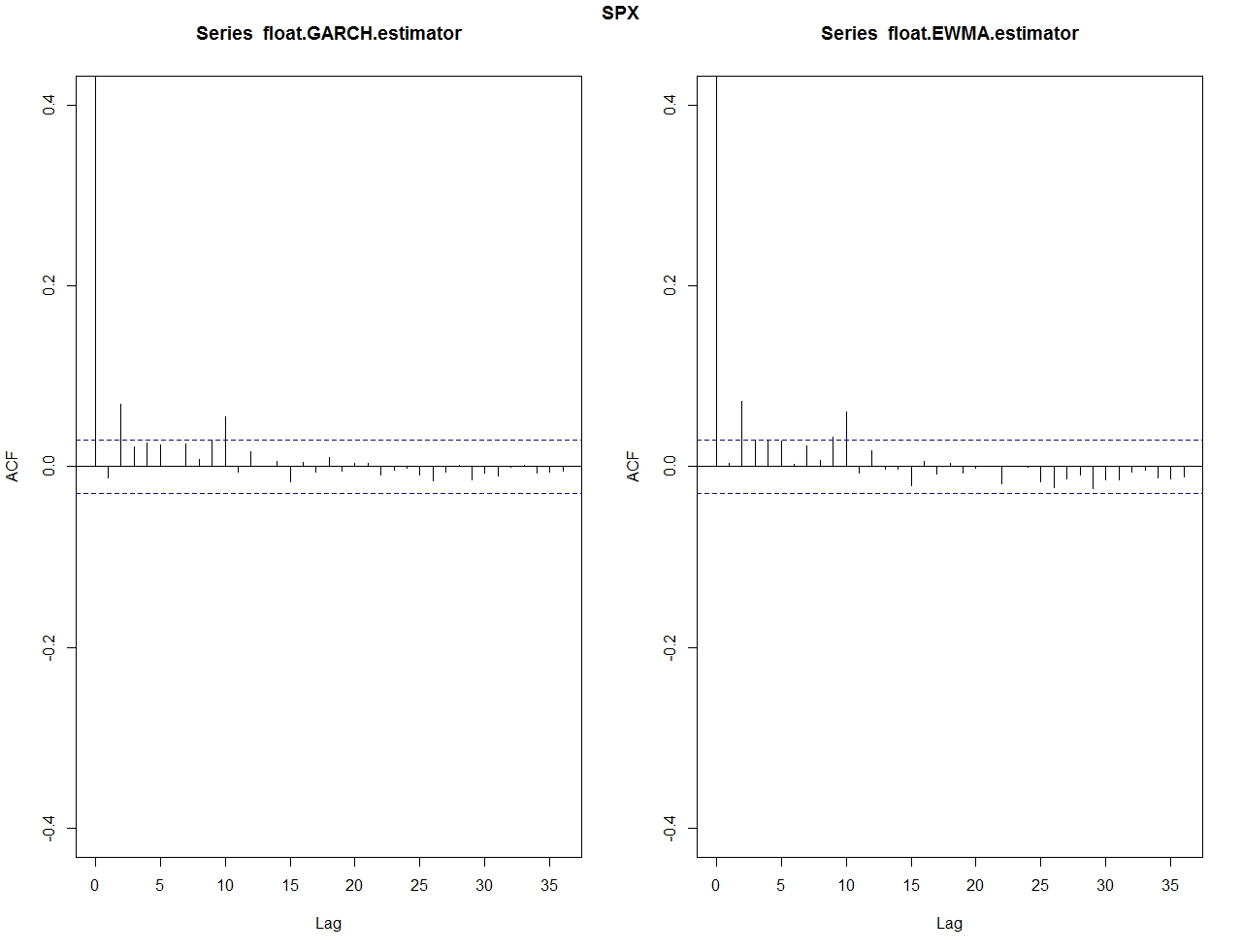}
  \caption{Autocorrelation test for the series of $(res_{t,i})^2$ in the SPX asset for GARCH and EWMA estimators with an increasing time window.}
  \label{float_comparison_acf_spx}
\end{figure}
In fact the ACF test in Figure \ref{float_comparison_acf_spx}  gives results basically similar to the ones of the original GARCH, so using a floating window on the estimation of the GARCH models is not a real gain, since it takes more computational time.
\subsection{Spurious effects in Moving Average estimation}
\begin{figure}[H]
  \includegraphics[width=\linewidth]{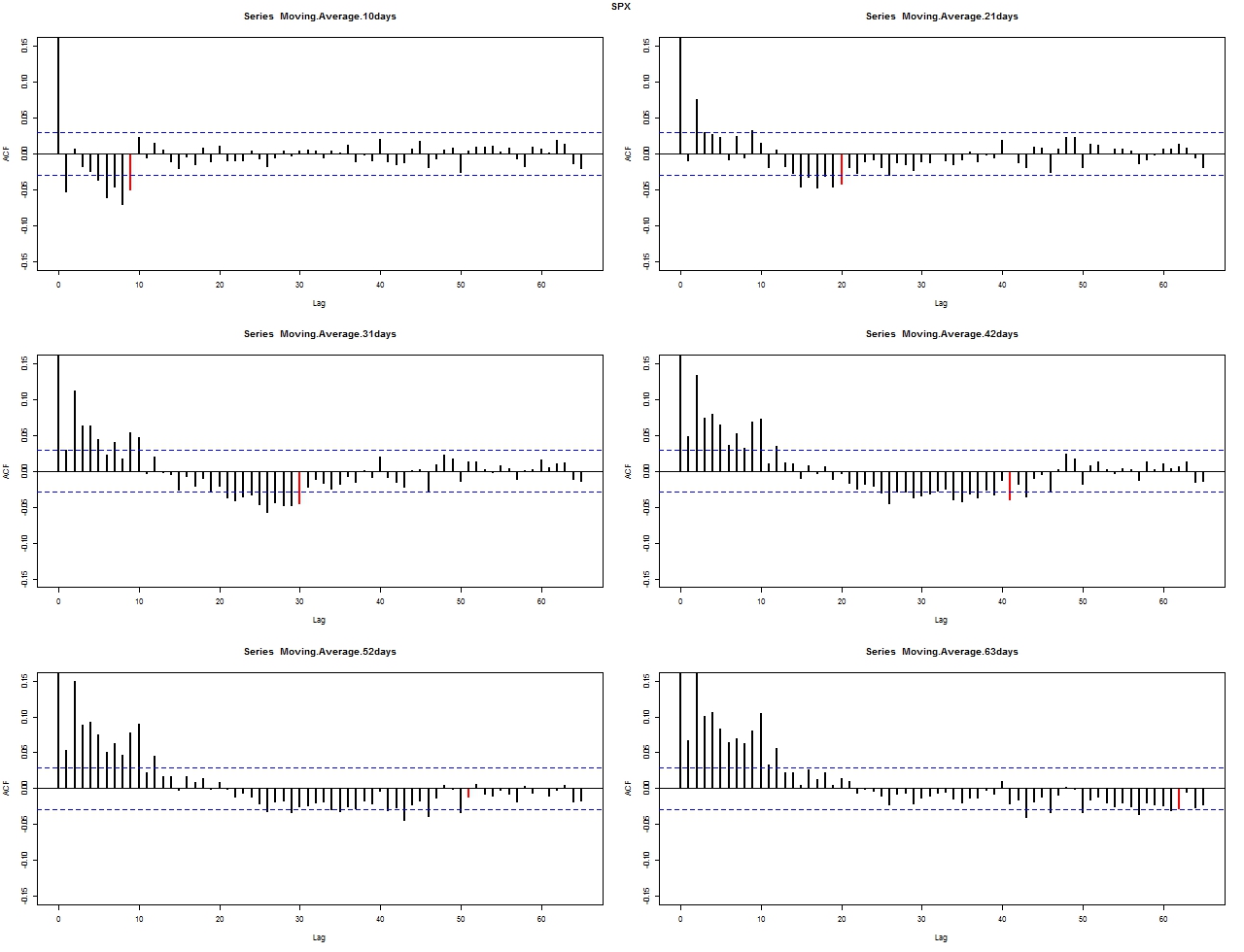}
  \caption{Autocorrelation test for the series of $(res_{t,i})^2$ in the SPX asset for Moving Average estimators with different window size.}
  \label{comp_moving_avarage_SPX}
\end{figure}

The first tests on autocorrelation of the Moving Averaged based estimation show that the residuals are far from being white noise. Another analysis can be made on the different sizes of the floating window used for the computation of the standard deviation.\\ 
We here used different sizes for the windows, and analyzed an interesting behavior of this estimator. \\
As we can see in the Figure \ref{comp_moving_avarage_SPX}, there is always negative correlation when the lag is exactly the size of the window. This is quite intuitive, since this lag corresponds to the fact that one return is going out from the time window, so it is not considered anymore by the estimator.\\
The red line, present in every graph, corresponds to the autocorrelation with lag exactly the size of the window of the Moving Average estimators.\\
As we can see, the red line presents always negative correlation, so this gives us an hint that the fact that it always corresponds to the day when we have new information coming, and old information exiting the window, this effect is just a spurious correlation element given by the method and not by the data.\\
\subsection{Conclusion}

After these tests we can recollect some interesting information:
\begin{itemize}
\item	\textbf{Constant volatility}: The poorest choice for the estimation of volatility. The Gaussianity of the residuals is given by the Gaussianity of the log returns, but the acf test and the Ljung - Box test prove that the squared residuals are far from a x squared. The volatility of the time series of the volatilities is of course 0, by construction. 
\item	\textbf{Intraday price volatility}: The good results on the ACF graphs and Ljung - Box test are balanced by very bad results in the QQplot. The volatility of the time series of volatilities is high, only the case of Yang Zhuang is better, because the statistical volatility is low by construction
\item	\textbf{Moving Average}: Not a good choice, since it lacks both from the normality point of view (especially in the squared of the residuals) and on the additional correlation studied in the acf comparison for different windows. 
\item	\textbf{Autoregressive models}: Good results in all the tests, even if in the ACF and Ljung - Box tests the results for intra day volatilities are better. The best results are for GARCH and EWMA, and these may be the best choice. GARCH because of its flexibility, EWMA because it depends only on one parameter.
\end{itemize}
\newpage
\phantom{bla}
\newpage
\section{Allocation strategies using Variance Swaps}
\subsection{Why derivatives on volatility}

Volatility swaps are forward contracts on future realized stock volatility. Variance swaps are similar contracts on variance, the square of future volatility. Both of these instruments provide an easy way for investors to gain exposure to the future level of volatility.\\
Unlike a stock option, whose volatility exposure is contaminated by its stock-price dependence, these swaps provide pure exposure to volatility alone. These instruments can be used to speculate on future volatility levels, to trade the spread between realized and implied volatility, or to hedge the volatility exposure of other positions or businesses.\\
Investors are increasingly looking at the derivatives space as a way for boosting returns of cross-asset portfolios, as depressed yields impact negatively the possibility of generating attractive returns in the current market.\\
Selling options is a popular strategy for extracting a positive carry, i.e., the return obtained from holding an asset,  out of the derivatives space, given the premium between implied and realized volatilities (and variances) on different markets.\\
The strategy displays attractive features over long-term periods, can be applied to several markets, and can deliver a positive return even in a low-rates environment, provided that the volatility premium remains wide enough.\\
The drawback of the strategy is that systematic options sellers can suffer large and sudden drawdowns during episodes characterized by spikes in volatility levels, which requires the ability of actively risk-managing the positions, ideally over short time-frames.\\
We will investigate a possibility for mitigating the losses associated with short volatility trades, which considers implementing long/short trades in the vol space. This implementation can allow removing the directional bias naturally associated with short-vol trades and the aforementioned large potential losses.
A comprehensive study of volatility and variance derivatives can be found in \cite{Derman} and \cite{Bossu}.

\subsection{Variance Swaps}
Although options market participants talk of volatility, it is variance, or volatility squared, that has more fundamental theoretical significance. This is so because the correct way to value a swap is to value the
portfolio that replicates it, and the swap that can be replicated most reliably (by portfolios of options of varying strikes, as we show later) is a variance swap.\\
A variance swap is a forward contract on annualized variance, the square of the realized volatility. Its payoff at expiration is equal to
\begin{equation}\label{varswap}
(\sigma_R^2 - K_{var} ) \times Notional,
\end{equation}
where $\sigma_R^2$ is the realized stock variance, computed as
$$
\sigma_R^2 = 100 \times \frac{252}{N} \times \displaystyle \sum_{t=1}^N \left( \ln \left( \frac{P_t}{P_{t-1}} \right) \right)^2,
$$
with $P_t$ the closing price at time $t$, $K_{var}$ is the delivery price for variance.\\
The holder of a variance swaps receive at expiration an amount of dollars equal to the $Notional$ multiplied by the points by which the realized variance of the stock $\sigma^2_R$ has exceeded the variance delivery price $K_{var}$.
\subsection{Replicating Variance Swaps}
Let us assume that the stock price evolution is given by
\begin{equation} \label{step1}
\frac{d S_t}{S_t} = \mu(t, \dots) dt + \sigma(t, \dots, ) d Z_t,
\end{equation}
where $Z_t$ is a standard Brownian motion and the drift $\mu$ and the continuously sampled volatility $\sigma$ are arbitrary functions of time and other parameters.\\
The theoretical definition of realized variance is the continuous integral
\begin{equation}\label{eqvarswap}
V = \frac{1}{T}\int_0^T \sigma^2(t, \dots) dt.
\end{equation}
This is a good approximation to the variance of daily returns used in the contract terms of most variance swaps.
The value of a forward contract $F$ on future realized variance with strike $K_{var}$ and notional equal to $1$ is the
expected present value of the future payoff in the risk-neutral world:
$$
F = \mathbb{E} \left[ e^{- r T} (V - K_{var}) \right].
$$
We will consider the Notional equal to $1$ in the whole analysis. \\ 
Here $r$ is the risk-free discount rate corresponding to the expiration
date $T$, and $\mathbb{E}[\cdot ]$ denotes the expectation under the risk neutral measure.\\
The fair delivery value of future realized variance is the strike $K_{var}$
for which the contract has zero present value:
\begin{equation}\label{eqKvar}
K_{var}= \mathbb{E} [V] =\frac{1}{T} \mathbb{E}  \left[ \int_0^T \sigma^2(t, \dots) dt \right] .
\end{equation}
We focus now on how to replicate the contract. We will focus after on how to value the contract in the case of stochastic volatility models.\\
By applying Ito's lemma to $\log (S_t)$ we get that (here we omit the dependence of $\mu$ and $\sigma$ from time and parameters)
\begin{equation}\label{step2}
d \log(S_t) = \left(\mu - \frac{\sigma^2}{2} \right) dt + \sigma dZ_t.
\end{equation}
If we subtract \eqref{step2} from \eqref{step1} we get
\begin{equation}\label{step3}
\frac{d S_t}{S_t} - d \log(S_t) = \frac{\sigma^2}{2}  dt.
\end{equation}
Integrating \eqref{step3} over time we can get the variance swap payoff:
\begin{equation} \label{step4}
\frac{2}{T} \int_0^T \frac{d S_t}{S_t} - \frac{2}{T}  \log\left( \frac{S_t}{S_0} \right) = \frac{1}{T} \int_0^T \sigma^2 dt.
\end{equation}
Taking the risk neutral expectation in \eqref{step4} we obtain
\begin{equation} \label{step5}
K_{var} = \frac{2}{T}  \mathbb{E} \left[ \int_0^T \frac{d S_t}{S_t}\right] - \frac{2}{T}  \mathbb{E} \left[ \log\left( \frac{S_T}{S_0} \right) \right].
\end{equation}
In a risk neutral world with a constant risk free rate $r$, the stock price evolves according to
$$
\frac{d S_t}{S_t} = r dt + \sigma(t, \dots, ) d Z_t,
$$
so we get that
\begin{equation}\label{part1}
\mathbb{E} \left[ \int_0^T \frac{d S_t}{S_t} \right] = rT + \mathbb{E} \left[ \int_0^T \sigma(t, \dots, ) d Z_t \right] = rT.
\end{equation}
We analyze the second part of the right hand side of \eqref{step5}. We can split the logarithm in two, in order to eliminate the part present in \eqref{part1}:
$$
\log\left( \frac{S_T}{S_0} \right) = \log\left( \frac{S_T}{S_0 e^{r T} } \right) + \log\left( \frac{S_0 e^{r T}}{S_0  } \right) = \log\left( \frac{S_T}{S_0 e^{r T} } \right) + rT,
$$
so that
\begin{equation}\label{part2}
K_{var} = \frac{2}{T}   \mathbb{E} \left[ - \log\left( \frac{S_T}{S_0 e^{r T}} \right) \right].
\end{equation}
For any twice differentiable payoff we have that (see e.g. \cite{CarrMadan}):
\begin{equation}\label{CM}
f(x) - f(y) = f'(y)(x - y) + \int_0^y f''(K)(K-x)^+ dK + \int_y^{+ \infty} f''(K) (x - K)^+ dK.
\end{equation}
Adapting the result in \eqref{CM} to our case, with $f(x) = -\log(x)$, $x=S_T$ and $y=S_0 e^{rT}$ we have that
\begin{align*}
-\log(S_T) + \log(S_0 e^{rT}) = & -\frac{S_T - S_0 e^{rT}}{S_0 e^{rT}} \\
&+ \int_0^{S_0 e^{rT}} \frac{1}{K^2} (K - S_T)^+ dK + \int_{S_0 e^{rT}}^{+ \infty} \frac{1}{K^2} (S_T - K)^+ dK,
\end{align*}
that can be rewritten as
\begin{align}\label{CM2}
-\log\left( \frac{S_T}{S_0 e^{rT}} \right) = & -\frac{S_T}{S_0 e^{rT}} + 1 \nonumber\\
&+ e^{rT} \int_0^{S_0 e^{rT}} \frac{1}{K^2} e^{- r T} (K - S_T)^+ dK\\
& + e^{rT} \int_{S_0 e^{rT}}^{+ \infty} \frac{1}{K^2} e^{- r T} (S_T - K)^+ dK. \nonumber
\end{align}
Taking the expected value on both the sides of \eqref{CM2} we get that
\begin{align*}
\mathbb{E} \left[ -\log\left( \frac{S_T}{S_0 e^{rT}} \right) \right] = &  -\frac{\mathbb{E} \left[ S_T \right]}{S_0 e^{rT}} + 1\\
& +  e^{rT} \mathbb{E} \left[ \int_0^{S_0 e^{rT}} \frac{1}{K^2} e^{- r T} (K - S_T)^+ dK \right] \\
& +  e^{rT} \mathbb{E} \left[ \int_{S_0 e^{rT}}^{+ \infty} \frac{1}{K^2} e^{- r T} (S_T - K)^+ dK \right].
\end{align*}
Using the fact that, under the risk neutral measure, $\mathbb{E} \left[ S_T \right] = S_0 e^rT$, and that we can use Fubini - Tonelli theorem to interchange integrals and expectations, since the integrand functions are positive, we obtain
\begin{equation}\label{CM3}
\mathbb{E} \left[ -\log\left( \frac{S_T}{S_0 e^{rT}} \right) \right] = e^{rT} \left( \int_0^{S_0 e^{rT}} \frac{1}{K^2} P_T(K) dK + \int_{S_0 e^{rT}}^{+ \infty} \frac{1}{K^2} C_T(K) dK \right),
\end{equation}
where $P(K)$ and $C(K)$ are, respectively, the price of a put and a call option with strike $K$ and maturity $T$.\\
We can finally write a replication formula for the variance swap:
\begin{equation}\label{CM4}
K_{var} = \frac{2}{T} e^{rT} \left( \int_0^{S_0 e^{rT}} \frac{1}{K^2} P_T(K) dK + \int_{S_0 e^{rT}}^{+ \infty} \frac{1}{K^2} C_T(K) dK \right).
\end{equation}
An important observation is that this replication formula assumes only that the spot price can be written in the form of \eqref{step1}. This approach is not compatible with jump diffusion processes, but it does not assume any particular model in a diffusion environment. A study of the impact of jumps in the replication formula can be found in \cite{Bossu}, but it is out of the scope of this study.
\subsubsection{Approximated replication and Corridor Variance Swap}
Due to the fact that it is impossible to have a continuous set of call and option prices, because of lack of out of the money liquidity and market standards, we can approximate the formula in \eqref{CM4} in order to adapt it to a discrete set of strikes for put and call prices. As we can see in \eqref{CM4}, the replicating formula includes only out of the money options. The approximation formula, which approximate the integral with a rectangle rule, is the following:
\begin{align}
K_{var} \approx  & \frac{2}{T} \frac{1}{S_0} \bigg( \sum_{i=1}^{N_{puts}} \frac{1}{(k_i^{put})^2} P_T(k_i^{put})  \left(k_i^{put} - k_{i-1}^{put}\right) \nonumber\\
&+ \sum_{i=1}^{N_{calls}} \frac{1}{(k_i^{call})^2} C_T(k_i^{call}) \left(k_i^{call} - k_{i-1}^{call}\right) \bigg),
\end{align}
where $P_T(k_i^{put})$ and $C_T(k_i^{call})$ are, respectively, the market put and call prices with maturity $T$ and the strikes $k_i^{put}$ and $k_i^{call}$ are in percentage of the forward value. In the equity space,  $k_i^{put} - k_{i-1}^{put}$ and $k_i^{call} - k_{i-1}^{call}$ are usually constant and equal to $5\%$.\\
This approximation assumes that $k_0^{put} = 0 $ and $k_{N_{calls}}^{call} = +\infty$, but since for such strikes the options are deeply out of the money, and so their value really close to $0$, we can assume that  $k_0^{put} = 50\% $ and $k_{N_{calls}}^{call} = 150 \%$ with a small (second) approximation error.\\
Motivated by the replication technique, it is possible to introduce a new type of derivative, that we present for completeness, but that we will not study in this work. The Corridor Variance Swap is a variance derivative similar to the variance swap, but it takes in account only a corridor of puts and calls in the replication formula. For example, if we are interested in a Corridor Variance Swap with corridor $\left[ 90\%, 110\% \right]$, (the most liquid derivatives in the market), the valuation formula would be:
\begin{align}
K^{corr}_{var} =  & \frac{2}{T} \frac{1}{S_0} \bigg( \sum_{\substack{i=1 \\ k_i^{put}  \geq 90\%} }^{N_{puts}} \frac{1}{(k_i^{put})^2} P_T(k_i^{put})  \left(k_i^{put} - k_{i-1}^{put}\right) \nonumber\\
&+ \sum_{\substack{i=1 \\ k_i^{call}  \leq 110\% }}^{N_{calls}} \frac{1}{(k_i^{call})^2} C_T(k_i^{call}) \left(k_i^{call} - k_{i-1}^{call}\right) \bigg),
\end{align}
\subsection{Variance Swaps under stochastic volatility}
The general strategy discussed in the previous section can be used to determine the fair variance and the hedging portfolio from the set of available options and their implied volatilities. Here we discuss how it is possible to find a pricing formula in different stochastic volatility models. This is an alternative way to the hedging approach developed in the previous section, since it gives only a pricing formula, without a flavor of the hedging strategy. Nevertheless, this approach is really important since it  permits to price Variance Swaps from the calibrated parameters of a model, and suggests the dependence of the price from the implied parameters.
All the formulas obtained are important also for the comparison between the market price, obtained using implied parameters, and the statistical price, obtained using parameters from the real world distribution. The formulas will allow a simple estimation of the statistical premium embedded in variance
swap prices.
\subsubsection{The SABR model}
The SABR (stochastic alpha, beta, rho) model, introduced in \cite{SABR},  is an extension of Black's model and of the CEV model, and assumes that the volatility follows the dynamic of a Geometric Brownian motion. The equations describing the model are
\begin{align}\label{SABRsde}
d S_t &= \sigma_t S_t^{\beta} d W^1_t \nonumber \\
d \sigma_t &=  \nu \sigma_t d W^2_t \\
\mathbb{E}\left[ dW_t^1 dW^2_t \right] &= \rho dt, \nonumber
\end{align}
where $S_t$ is the asset price, $\sigma_t$ is the volatility, $\nu$ is the volvol parameter, $\rho$ the correlation between the two Brownian motions $W_t^1$ and $W_t^2$, and initial values $S_0 = S_0$ and $\sigma_0 = \alpha$.\\
As described in \eqref{eqKvar}, the fair value of a Variance Swap can be computed as
$$
K_{var} = \frac{1}{T} \mathbb{E} \left[ \int_0^T \sigma^2_t dt  \right].
$$
Since the volatility follows the dynamic of a GBM, it is straightforward the computation of the solution:
\begin{equation*}
\sigma_t = \alpha e^{ \nu W_t - \frac{\nu^2}{2}t}.
\end{equation*}
The computation of the Variance Swap is then
\begin{align*}
K_{var}^{SABR} & = \frac{1}{T} \mathbb{E} \left[  \int_0^T \alpha^2 e^{ 2 \nu W_t - \nu^2 t}  dt \right] \\
& = \frac{\alpha^2} {T} \mathbb{E} \left[   \int_0^T e^{ 2 \nu W_t - \frac{ (2 \nu)^2}{2} t} e^{\nu^2 t} dt  \right] \\
& = \frac{\alpha^2} {T}    \int_0^T e^{\nu^2 t} \mathbb{E} \left[e^{ 2 \nu W_t - \frac{ (2 \nu)^2}{2} t} \right]  dt  \\
& = \frac{\alpha^2} {T}    \int_0^T e^{\nu^2 t}  dt,
\end{align*}
where we used Fubini Tonelli theorem since the integrand is positive and the fact that $e^{ 2 \nu W_t - \frac{ (2 \nu)^2}{2} t}$ is a martingale. We obtain that
\begin{equation}\label{KvarSABR}
K_{var}^{SABR} = \frac{\alpha^2} {\nu^2 T}\left( e^{\nu^2 T} - 1 \right).
\end{equation}

\subsubsection{The Heston model}
The Heston model, introduced in \cite{Heston}, is one of the most famous stochastic volatility models, and assumes that the variance follows the dynamic of a Cox - Ingersoll - Ross process:
\begin{align}\label{Hestonsde}
d S_t &= r S_t dt + \sigma_t S_t d W^1_t \nonumber \\
d \sigma^2_t &= \kappa( \theta - \sigma_t^2) dt +  \nu \sigma_t d W^2_t \\
\mathbb{E}\left[ dW_t^1 dW^2_t \right] &= \rho dt, \nonumber
\end{align}
where $S_t$ is the asset price, $\sigma^2_t$ is the variance, $\kappa$ is the variance mean reversion parameter, $\theta$ is the long term variance, $\nu$ is the volvol parameter, $\rho$ the correlation between the two Brownian motions $W_1^t$ and $W_2^t$, and initial values $S_0 = S_0$ and $\sigma^2_0 = \sigma^2_0$.\\
The solution of the CIR process is more tricky to obtain, but can still be computed. Applying Ito Lemma to $e^{\kappa t}\sigma^2_t$ we get
\begin{align*}
d e^{\kappa t}\sigma^2_t & = e^{\kappa t} \kappa \sigma^2_t dt + e^{\kappa t} \kappa \theta dt - e^{\kappa t} \kappa \sigma^2_t dt + e^{\kappa t} \nu \sigma_t^2 dW^2_t\\
& = e^{\kappa t} \kappa \theta dt  + e^{\kappa t} \nu \sigma_t dW^2_t,
\end{align*}
so we have that
\begin{align*}
e^{\kappa t}\sigma^2_t & = \sigma^2_0 + \int_0^t e^{\kappa s} \kappa \theta ds + \int_0^t e^{\kappa s} \nu \sigma_s dW^2_s \\
& = \sigma^2_0 + \theta \left(e^{\kappa t} -1\right) + \int_0^t e^{\kappa s} \nu \sigma_s dW^2_s,
\end{align*}
and the solution is
\begin{equation}\label{cir}
\sigma^2_t = \theta + e^{- \kappa t}\left(\sigma^2_0 - \theta \right) + \int_0^t e^{\kappa (s - t)} \nu \sigma_s dW^2_s.
\end{equation}
We can then compute the value of the Variance Swap:
\begin{align*}
K_{var}^{Heston} & = \frac{1}{T} \mathbb{E} \left[  \int_0^T \left(\theta + e^{- \kappa t}\left(\sigma^2_0 - \theta \right) + \int_0^t e^{\kappa (s - t)} \nu \sigma_s dW^2_s\right)  dt \right] \\
& = \frac{1}{T}   \int_0^T \mathbb{E} \left[ \theta + e^{- \kappa t}\left(\sigma^2_0 - \theta \right) \right] dt + \int_0^T \mathbb{E}\left[\int_0^t e^{\kappa (s - t)} \nu \sigma_s dW^2_s \right]  dt  \\
& = \theta + \frac{\sigma^2_0 - \theta}{T} \int_0^T e^{- \kappa t} dt,
\end{align*}
where we used Fubini Tonelli since the integrand are positive, and we assumed a regularity in the process $\sigma_t$ in order to have the second expectation equal to $0$.
We obtain that
\begin{equation}\label{KvarHeston}
K_{var}^{Heston} = \theta + \frac{\sigma^2_0 - \theta}{\kappa T} \left(1 -  e^{- \kappa T} \right).
\end{equation}
\subsubsection{The Stein Stein model}
The Stein Stein model, introduced in \cite{SteinStein}, is historically the first stochastic volatility model that presents correlation between the price and the volatility. The volatility follows the dynamic of an Ornstein - Uhlenbeck process:
\begin{align}\label{Steinsde}
d S_t &= r S_t dt + \sigma_t S_t d W^1_t \nonumber \\
d \sigma_t &= \kappa( \theta - \sigma_t) dt +  \nu d W^2_t \\
\mathbb{E}\left[ dW_t^1 dW^2_t \right] &= \rho dt, \nonumber
\end{align}
where $S_t$ is the asset price, $\sigma_t$ is the volatility, $\kappa$ is the volatility mean reversion parameter, $\theta$ is the long term volatility, $\nu$ is the volvol parameter, $\rho$ the correlation between the two Brownian motions $W_1^t$ and $W_2^t$, and initial values $S_0 = S_0$ and $\sigma_0 = \sigma_0$.\\
The solution of the Ornstein - Uhlenbeck process can be obtained similarly to the one of the CIR process in the previous section. We still need to apply the Ito Lemma to $e^{\kappa t}\sigma_t$ :
\begin{align*}
d e^{\kappa t}\sigma_t & = e^{\kappa t} \kappa \sigma_t dt + e^{\kappa t} \kappa \theta dt - e^{\kappa t} \kappa \sigma_t dt + e^{\kappa t} \nu dW^2_t\\
& = e^{\kappa t} \kappa \theta dt  + e^{\kappa t} \nu d W^2_t,
\end{align*}
so we have that
\begin{align*}
e^{\kappa t}\sigma_t & = \sigma_0 + \int_0^t e^{\kappa s} \kappa \theta ds + \int_0^t e^{\kappa s} \nu dW^2_s \\
& = \sigma_0 + \theta \left(e^{\kappa t} -1\right) + \nu \int_0^t e^{\kappa s} \sigma_s dW^2_s,
\end{align*}
so the solution is
\begin{equation}\label{ou}
\sigma_t = \theta + e^{- \kappa t}\left(\sigma_0 - \theta \right) + \nu \int_0^t e^{\kappa (s - t)}  dW^2_s.
\end{equation}
In the computation of the Variance Swap, we will need the expected value of the variance, so we make the computations in a separated part:
\begin{align*}
\mathbb{E}\left[ \sigma_t^2 \right]  = & \left( \theta + e^{- \kappa t}\left(\sigma_0 - \theta \right) \right)^2 + 2 \nu \left( \theta + e^{- \kappa t}\left(\sigma_0 - \theta \right) \right) \mathbb{E}\left[ \int_0^t e^{\kappa (s - t)}  dW^2_s \right] \\
&  + \nu^2 \mathbb{E}\left[ \left(\int_0^t e^{\kappa (s - t)}  dW^2_s \right)^2 \right] \\
 = & \theta^2 + e^{- 2 \kappa t}\left(\sigma_0 - \theta \right)^2 + 2 \theta e^{- \kappa t}\left(\sigma_0 - \theta \right) + \nu^2 \int_0^t e^{2 \kappa (s - t)}  ds,
 \end{align*}
 where we used Fubini Tonelli as usual, and the Ito isometry. We have that
 \begin{equation}\label{expou}
\mathbb{E}\left[ \sigma_t^2 \right] = \theta^2 + e^{- 2 \kappa t}\left(\sigma_0 - \theta \right)^2 + 2 \theta e^{- \kappa t}\left(\sigma_0 - \theta \right) + \nu^2 \frac{1 - e^{- 2 \kappa t}}{2 \kappa}.   
\end{equation}
We can then compute the value of the Variance Swap:
\begin{align*}
K_{var}^{Stein} & = \frac{1}{T}  \int_0^T \left(\theta^2 + e^{- 2 \kappa t}\left(\sigma_0 - \theta \right)^2 + 2 \theta e^{- \kappa t}\left(\sigma_0 - \theta \right) + \nu^2 \frac{1 - e^{- 2 \kappa t}}{2 \kappa} \right)  dt \\
& = \theta^2 + \frac{\nu^2}{2 \kappa} + 2 \theta \left(\sigma_0 - \theta \right) \frac{1}{T} \int_0^T  e^{- \kappa t}  dt + \left( \left(\sigma_0 - \theta \right)^2 - \frac{1}{2 \kappa} \right) \frac{1}{T} \int_0^T  e^{-2 \kappa t}  dt,
\end{align*}
and we get that
\begin{equation}\label{KvarStein}
K_{var}^{Stein} = \theta^2 + \frac{\nu^2}{2 \kappa} + 2 \theta \left(\sigma_0 - \theta \right) \frac{ 1 - e^{- \kappa T}}{\kappa T} + \left( \left(\sigma_0 - \theta \right)^2 - \frac{\nu^2}{2 \kappa} \right) \frac{ 1 - e^{- 2 \kappa T}}{2 \kappa T}.
\end{equation}
\subsubsection{The $\lambda$-SABR model}
The $\lambda$-SABR model, introduced in \cite{lambdaSABR}, extends the SABR model of \cite{SABR}, introducing mean reversion in the volatility, which follows the dynamic of a modified Ornstein - Uhlenbeck process:
\begin{align}\label{lambdaSABRsde}
d S_t &= \sigma_t S_t^{\beta} d W^1_t \nonumber \\
d \sigma_t &= \kappa(\theta - \sigma_t)dt + \nu \sigma_t d W^2_t \\
\mathbb{E}\left[ dW_t^1 dW^2_t \right] &= \rho dt, \nonumber
\end{align}
where $S_t$ is the asset price, $\sigma_t$ is the volatility, $\kappa$ is the volatility mean reversion parameter, $\theta$ is the long term volatility, $\nu$ is the volvol parameter, $\rho$ the correlation between the two Brownian motions $W_1^t$ and $W_2^t$, and initial values $S_0 = S_0$ and $\sigma_0 = \alpha$.\\
The computations of the expected value of the variance and of the Variance Swap are long and tedious, so we will present only the final result:
\begin{align}
K_{var}^{\lambda - SABR}  = &\left(\frac{\kappa \theta}{\kappa + \frac{\nu}{2}}\right)^2 + \frac{\alpha^2}{(2 \kappa - \nu^2)T} \left( 1 - e^{- (2 \kappa - \nu^2)T} \right) \nonumber\\
& + \frac{ \kappa^2 \theta^2}{2 \left( \kappa + \frac{\nu^2}{2} \right)^3 T} \left( 1 - e^{ -2 \left( \kappa + \frac{nu^2}{2} \right) T }\right) + \frac{2 \alpha \theta}{\kappa \left(\kappa + \frac{\nu^2}{2} \right) T} \left( 1 - e^{- \kappa T} \right) \nonumber\\
& + \frac{2 \alpha \kappa \theta}{ \left(\kappa + \frac{\nu^2}{2} \right)  \left(2 \kappa + \frac{\nu^2}{2} \right) T} \left( 1 - e^{ - \left(2 \kappa + \frac{\nu^2}{2} \right) T} \right).
\end{align}
\subsection{A brief comparison of the pricing methods}
A first test on the goodness of the pricing techniques is made on the SPX 1 month Variance Swaps. Since the SABR model is the easiest and most parsimonious model to calibrate, we will use it as a proxy for the pricing via stochastic volatility models, and we will also analyze the pricing via replication.\\
In the SABR pricing formula we calibrate a SABR model with $\beta=1$ and we use the parameters that we need, i.e., $\alpha$ and $\nu$.
The calibration routine computes different parameters for different maturities, from $1$ week to $1$ year. In the example we will focus on maturity $1$ month. The period goes from March 2006 to March 2016, and we compare one variance swap per month, so $12$ per year.
\begin{figure}[H]
  \includegraphics[width=\linewidth]{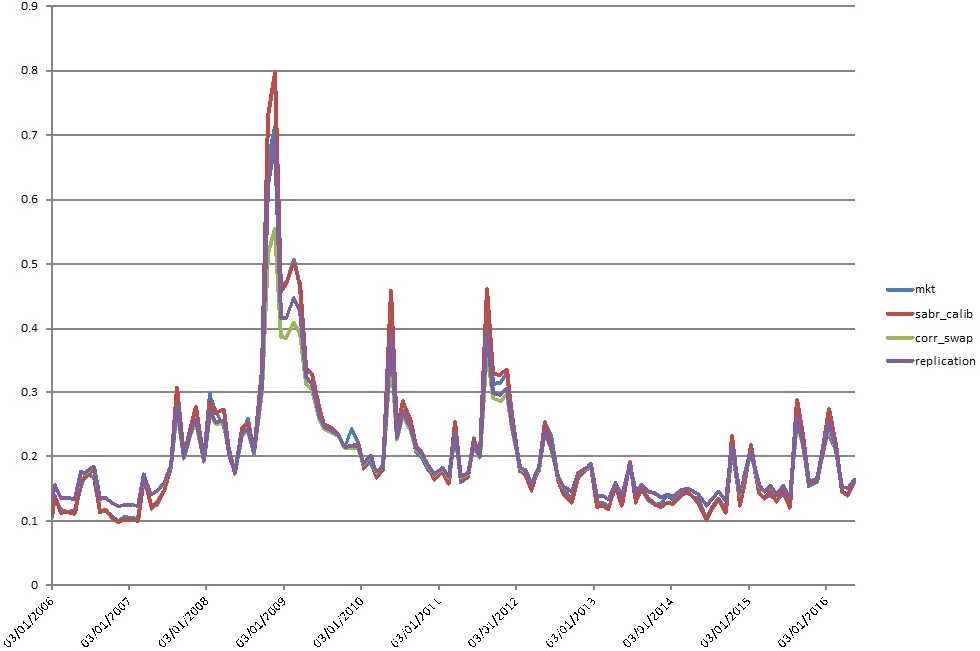}
  \caption{Pricing of Variance Swap via replication and via the SABR formula, comparison with market data.}
  \label{comparison_pricing_SPX}
\end{figure}
As we can see, the SABR formula gives an almost perfect fit, where the differences are probably given by the calibration routine. In fact, the SABR model is mostly used for FX and rates derivatives, while other models are used for the Equity space. The objective of this numerical study was not to find the best calibration of the market smile, but just a first proxy for the pricing of Variance Swap. The difference is pricing is mostly due to the incapacity of the SABR smile to fit perfectly the market smile, mostly for deep out of the money options. Nevertheless, we are quite happy of the result, as the SABR formula can be used as a good pricing estimation for Variance Swap.\\ 
On the other side, the results given by the replication formula for the corridor swap are quite poor but the motivation is easy to understand. The first error is given by the fact that we approximate an integral with a finite sum, and that is an error that will always be done, unless we try to bootstrap the missing strikes of the replication formula. Since we wanted to use only market data, we did not fit the market smile with artificial values. In addition, we only used strikes from $90\%$ to $110\%$ of the moneyness. This is due to the fact that Equity puts and calls are not as liquid as the FX vanilla options, so for deep out of the money options the bid-ask spread was too big to be considered. As we can see the biggest difference is for the peak in volatility of August 2008, while for the others values the pricing is still a good proxy.\\
The replication price is obtained using the replication formula, and keeping constant volatility for deep out of the money options. Practically, we keep the volatility constant to the value of $90\%$ of the moneyness for the strikes from $50\%$ to $90\%$, and the value of $110\%$ for the strikes from $110\%$ to $150\%$. Even if this choice of constant volatility outside the range $[90\%, 110\%]$ is quite poor and does not respect non arbitrage conditions, the pricing formula seems fitting the values of the market data.
\subsection{Mark to Market price}
Since Variance Swaps can be exercised only at maturity, with the final payoff given by the total realized variance minus the strike fixed at the beginning of the contract, it is important, in terms of P\&L of strategies using Variance Swaps, to consider also the Mark to Market value of this derivative. We know that the Mark to Market value of a Variance Swap at the beginning of the contract is of course $0$, while at maturity $T$, counting $N$ days, is (in the case we are long)
$$
100 \times \frac{252}{N} \times \sum_{i=1}^{N} \ln\left( \frac{P_i}{P_{i-1}} \right)^2 - K_{var},
$$
and we remember that, as usual, 
$$
 K_{var}  = 100 \times \frac{252}{N} \times \sum_{i=1}^{N} \mathbb{E} \left[ \ln\left( \frac{P_i}{P_{i-1}} \right)^2 \right].
$$
Let us call $imp_t^T$ the value of a (synthetic) Variance Swap at time $t$ with maturity $T$, so such that
\begin{equation}
 imp_t^T  = 100 \times \frac{1}{(T-t)} \times \mathbb{E} \left[ \sum_{i=1+252 \times t}^{252 \times T} \ln\left( \frac{P_i}{P_{i-1}} \right)^2 \right].
\end{equation}
It can be easily seen that $K_{var} = imp_0^T$.\\
The Mark to Market value of a Variance Swap contract is then , assuming the interest rates equal to $0$, a convex combination of the realized variance up to time $t$ and the predicted variance from time $t$, minus the Variance Swap strike :  
\begin{equation}
MtM_t = \frac{t}{T} \sum_{i=1}^{252 \times t} \ln\left( \frac{P_i}{P_{i-1}} \right)^2 + \frac{T-t}{T}  imp_t^T - K_{var},
\end{equation}
where we assume that $\sum_{i=1}^{0} \cdot = 0$. We can easily see that this definition is what we are looking for, in the case of  $t=0$ and $t=T$, since
\begin{align*}
MtM_0 & =  \frac{0}{T} \sum_{i=1}^{252 \times 0} \ln\left( \frac{P_i}{P_{i-1}} \right)^2 + \frac{T}{T}  imp_0^T - K_{var} \\
& = imp_0^T - K_{var} = 0,
\end{align*}
and
\begin{align*}
MtM_T & = \frac{T}{T} \sum_{i=1}^{252 \times T} \ln\left( \frac{P_i}{P_{i-1}} \right)^2 + \frac{T-T}{T}  imp_t^T - K_{var} \\
& = \sum_{i=1}^{252 \times T} \ln\left( \frac{P_i}{P_{i-1}} \right)^2 - K_{var}.
\end{align*}
The main issue is how to compute the element $imp_t^T$. Since, as we have seen in the previous Section, the pricing using the SABR formula is a good approximation of the market price, we will use that approach, specifically
\begin{equation}
imp_t^T = \frac{\alpha^2} {\nu^2 (T-t)}\left( e^{\nu^2 (T-t)} - 1 \right),
\end{equation}
where $\alpha$ and $\nu$ are the parameters of the SABR model computed at the beginning of the contract.
The dynamic of the Mark to Market price can be seen in the following Figure:
\begin{figure}[H]
  \includegraphics[width=\linewidth]{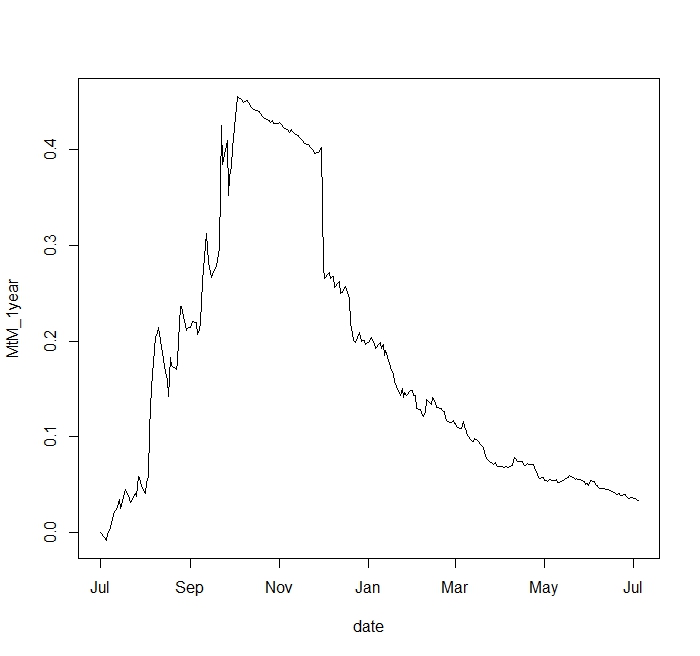}
  \caption{Mark to Market value of a $1$ year Variance Swap on the SPX index. The starting day is July 1st, 2011.}
  \label{MtM_1year}
\end{figure}
The Mark to Market value in Figure \ref{MtM_1year} is scaled by a factor $\frac{1}{K_{var}}$. This choice, suggested in \cite{Bossu}, is done in order to have the vega of the Mark to Market value constant.\\
As we can see, if we did not consider all the behavior of the Mark to Market price from the beginning to maturity, we would have considered only the initial value $0$ and the final value $0.0332$. Meanwhile, the evolution of the price has been consistently different. This suggests that keeping the contract up to maturity is not always a good idea, since, for example, selling it in November, would have given a much better return to the strategy. Even if we do not consider the possibility to buy or sell the contract before maturity, it is important to consider what could have been the value of the contract, in the view of P\&L calculations.
\subsection{Back test on a long strategy}
In this section we will focus on a simple strategy, that we will apply for all the maturities considered, i.e., $1$ month, $3$ months, $6$ months and $1$ year. The idea behind this strategy is that we enter in a new contract every month, and we keep every derivative up to maturity. We are keeping the variance swaps until maturity, and at the end of the contract we receive the difference between the realized variance and the strike fixed at the beginning of the contract.\\ 
In the case of $1$ month Variance Swaps we are keeping only one derivative in our portfolio at every time $t$, while, for example in the case of $3$ months Variance Swaps, we start with one derivative, then after one month we add a second Variance Swap, and at the second month we add the third Variance Swap. At the third month the first Variance Swap is expired, and we can add a new one. In this way, we have always at maximum $3$ Variance Swaps in the portfolio, in the case of $3$ months Variance Swaps. \\ 
Since the longest maturity considered is $12$ months, the maximum number of Variance Swaps in the portfolio would be $12$. This is done from July 2011 to June 2016.
\begin{figure}[H]
  \includegraphics[width=\linewidth]{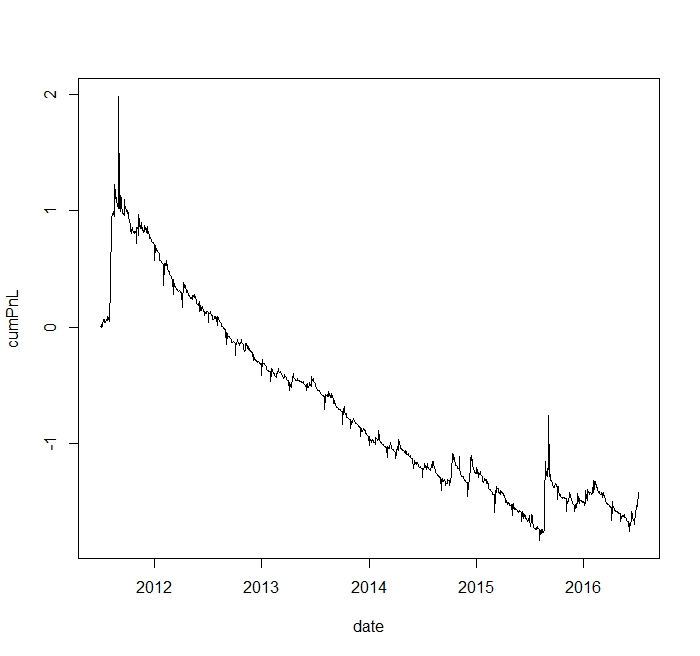}
  \caption{Profit and Loss of the long strategy using  $1$ month Variance Swap on the SPX index. The starting day is July 1st, 2011.}
  \label{PnL_1month}
\end{figure}
\begin{figure}[H]
  \includegraphics[width=\linewidth]{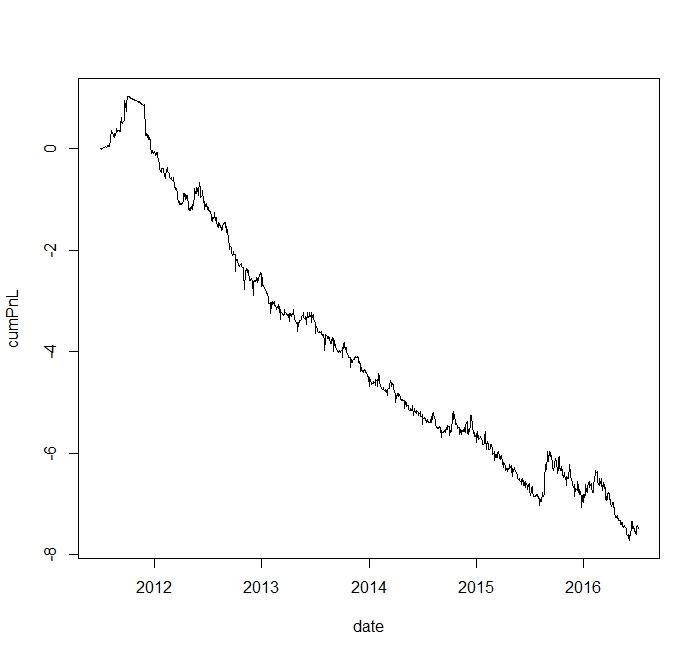}
  \caption{Profit and Loss of the long strategy using  $1$ year Variance Swap on the SPX index. The starting day is July 1st, 2011.}
  \label{PnL_1year}
\end{figure}
Figures \ref{PnL_1month} and Figure \ref{PnL_1year} show similar trend but different peculiarities. In fact, the spikes presents in the $1$ month strategy are more important then the ones of the $1$ year strategy. The $1$ year strategy presents a more interesting trend, with less volatility with respect to the monthly case. This can also be shown in Table \ref{PnL}.

The final Sharpe ratios are calculated with the following formula: we call $cumMtM_t$ the cumulative Mark to Market value of the strategy, i.e., the sum of the Mark to Market values of the Variance Swaps which are in the portfolio at time $t$, so we are considering a maximum of $12$ Variance Swaps. Let $retMtM_t$ the returns of $cumMtM_t$, 
$$
retMtM_t = cumMtM_t - cumMtM_{t-1}.
$$
The Sharpe ratios of this strategy are then computed as the mean of the time series $retMtM_t$ divided by its standard deviation.\\
This computation has be done for all the maturities and for five different assets, two equities and three FXs.
\begin{table}[H]
\centering
\begin{tabular}{c |c |c |c |c}
          & 1 month & 3 months & 6 months & 1 year \\
          \hline
SPX       & -0.31   & -0.77    & -1.10    & -1.34  \\
EUROSTOXX & -1.01   & -0.13    & -1.03    & -      \\
EURUSD    & -0.09   & -0.48    & -0.44    & -0.87  \\
GBPUSD    & -0.48   & -0.51    & -0.57    & -0.31  \\
USDJPY    & -0.26   & -0.31    & -0.34    & -0.68 
\end{tabular}
\caption{Sharpe Ratio of the only long strategies on different assets, using Variance Swaps with different maturities.}
\label{PnL}
\end{table}
\subsection{Stationarity test on implied volatilities and Variance swaps}
In this section we will be interested in highlighting mean reverting features appearing on cross-asset
derivatives. A full cross-sectional analysis, focusing on long/short trades, will come in
handy for magnifying the mean-reversion effects already appearing on single assets: for this,
we will focus on Equity and FX spaces, although other asset classes could be easily taken into
account as well.
Mean reversion is an important characteristic of a time series. From a strategic point of view, knowing the mean reversion time of a price series can give suggestions on when to buy / sell, in order to amplify the final return of the strategy.
In order to study mean reversion of a time series, we introduce the basic concepts.\\
An Autoregressive $1$ model, also known as $AR (1)$ specifies that the variable $y_t$ depends linearly from its previous value and on a stochastic term:
\begin{equation}\label{AR}
y_t - m = A ( y_{t-1} - m) + \varepsilon_t.
\end{equation}
The process is stationary if $|A| < 1$, and in this case it is possible to compute the mean of the process, that is $\mathbb{E}\left[ y_t \right] = m$.  A common way to quantify the speed of mean reversion of an autoregressive process is the half-life, which is defined as the working days needed so that the magnitude of the forecast become half of that of the forecast origin. Since, given a starting value $y_k$, the forecast value after $n$ working days is $\hat{y}_{n+k} = A^k y_n$, we have that $A^k y_n = \displaystyle \frac{1}{2} y_n$, so that the half life can be computed as
\begin{equation}
k = \frac{\log\left(\frac{1}{2} \right) }{\log \left( |A| \right) }. 
\end{equation}
If the parameter $A$ is close to $0$, then the half life of the process is really small, and it is possible to exploit this property from an allocation point of view.\\
\subsubsection{Correlation between implied volatilities and Variance Swap}
Since Variance Swap are not really liquid on all asset classes, and due to the fact that the standard version of Bloomberg does not display the time series of Variance Swaps, it is useful to find a good proxy for the price of Variance Swaps, that can give the same behavior of the time series. Following the formulas of the stochastic volatility models, especially in the SABR case, we can see that the prices are always given by the parameter $\sigma_0$ multiplied by a corrective factor. So a good idea would be to study the correlation between the time series of the ATM volatility and of the Variance Swap.
\begin{figure}[H]
  \includegraphics[width=\linewidth]{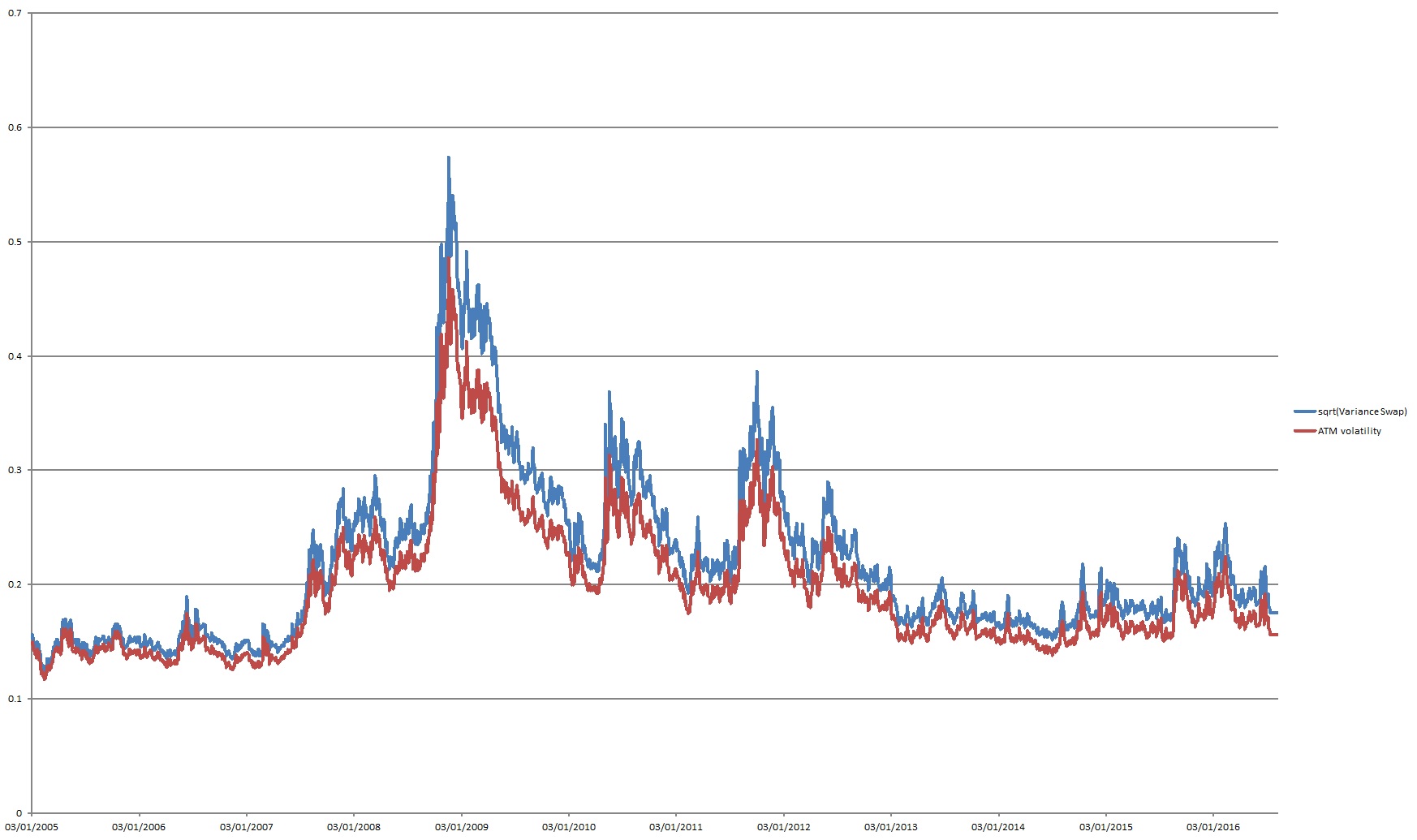}
  \caption{Time series of the square root of the $1$ year Variance Swap and $1$ year At The Money Volatility on the SPX index. The Variance Swap prices are obtained using the replication formula.}
  \label{varswap_volatm_1year_SPX}
\end{figure}
Motivated by the result in Figure \ref{varswap_volatm_1year_SPX}, we analize the correlation between Variance Swap prices and ATM volatility:
\begin{table}[H]
\centering
\begin{tabular}{c|c|c|c|c|c}
          & 1 m &  3 m & 6 m & 12 m  & 18 m \\
\hline
SPX       & 99.82\% &  99.74\%  & 99.79\%  & 99.94\% & 99.89\%   \\
Eurostoxx & 99.15\% &  99.38\%  & 99.12\%  & 98.97\% & 98.25\%   \\
UKX       & 99.52\% &  99.51\%  & 99.41\%  & 99.17\% & 98.67\%   \\
Nikkei    & 98.24\% &  98.00\%  & 97.32\%  & 97.38\% & 97.36\%  
\end{tabular}
\caption{Correlation between the time series of the square root of the Variance Swap and the At The Money Volatility.}
\label{VSvsatm_correlation}
\end{table}
As we can see in Table \ref{VSvsatm_correlation}, it is natural to study directly the behavior of the ATM Volatility, since it will give all the properties needed from the Autoregressive model, due to the fact that the correlation between ATM volatility and (the square root of) the price of Variance Swaps is always greater than $97\%$, on all assets and all maturities.

\subsubsection{Stationarity test on implied volatilities: the Dickey - Fuller test }
Stationarity is a statistical concept which refers to the property of a financial variable to oscillate in a range around a pre-specified value. If that is the case, one can set up a mean reversion strategy for trading the dislocations of an asset from its convergence value.\\
The concept of stationarity is directly related to that of unit roots in the context of autoregressive models. A unit-root essentially mimics a Brownian motion-type of behavior, whereby the latest price encompasses all the relevant information, so that there is no corrective force towards any other pre-determined value than the latest value itself.\\
Stationarity of financial assets can be assessed by performing a Dickey Fuller test. The test investigates the unit root property of the time series in the framework of autoregressive models, by testing a set of autocorrelation (ACF) functions of the time series.\\
The null hypothesis of the test is that a unit root is present in the time series and that the asset is non-stationary. The result of the test is a p-value related to the probability of rejecting the null hypothesis when the latter is in fact true. In practice one fixes a threshold or significance level (usually at 5\%) and rejects the null-hypothesis if the p-value is below the threshold: essentially, when the p-value of the Dickey-Fuller test applied on an asset is below the threshold, we reject the hypothesis that that asset is non-stationary, which supports the suggestion that the asset might be in fact stationary. When the p-value is above the threshold, we simply do not have enough elements for rejecting the null hypothesis of non-stationarity.\\
The test can be applied to any financial asset; in this piece we have applied to a set of spot, return and volatility time series. For numerical tests, we have relied on the \textit{adf.test} function from the \textit{tseries} \textbf{R} package.
\begin{table}[H]
\centering
\begin{tabular}{c | c c c c c}
          & 1m                           & 3m                           & 6m                           & 12m                          & 18m  \\
          \hline
SPX       & 0.08                         & 0.25                         & 0.36                         & 0.43                         & 0.46 \\
Eurostoxx & \cellcolor[HTML]{34FF34}0.01 & \cellcolor[HTML]{34FF34}0.05 & 0.10                         & 0.21                         & 0.27 \\
Footsie   & \cellcolor[HTML]{34FF34}0.01 & \cellcolor[HTML]{34FF34}0.03 & \cellcolor[HTML]{34FF34}0.05 & 0.06                         & 0.06 \\
Nikkei    & \cellcolor[HTML]{34FF34}0.01 & \cellcolor[HTML]{34FF34}0.04 & 0.10                         & \cellcolor[HTML]{34FF34}0.04 & 0.17 \\
Swiss     & \cellcolor[HTML]{34FF34}0.01 & 0.07                         & 0.21                         & 0.34                         & 0.39 \\
Canada    & \cellcolor[HTML]{34FF34}0.03 & 0.17                         & 0.34                         & 0.40                         & 0.40 \\
Australia & 0.09                         & 0.27                         & 0.26                         & 0.46                         & 0.36 \\
Norway    & \cellcolor[HTML]{34FF34}0.03 & \cellcolor[HTML]{34FF34}0.03 & 0.40                         & 0.34                         & 0.28 \\
Sweden    & \cellcolor[HTML]{34FF34}0.02 & 0.09                         & 0.15                         & 0.11                         & 0.10 \\
EURUSD    & \cellcolor[HTML]{34FF34}0.02 & 0.13                         & 0.28                         & 0.39                         & 0.67 \\
GBPUSD    & \cellcolor[HTML]{34FF34}0.01 & 0.09                         & 0.23                         & 0.28                         & 0.64 \\
USDJPY    & \cellcolor[HTML]{34FF34}0.01 & \cellcolor[HTML]{34FF34}0.01 & \cellcolor[HTML]{34FF34}0.03 & 0.06                         & 0.43 \\
USDCHF    & \cellcolor[HTML]{34FF34}0.01 & \cellcolor[HTML]{34FF34}0.02 & 0.08                         & 0.19                         & 0.50 \\
USDCAD    & \cellcolor[HTML]{34FF34}0.03 & 0.14                         & 0.25                         & 0.31                         & 0.46 \\
AUDUSD    & \cellcolor[HTML]{34FF34}0.01 & 0.06                         & 0.10                         & 0.26                         & 0.55 \\
NZDUSD    & \cellcolor[HTML]{34FF34}0.01 & \cellcolor[HTML]{34FF34}0.05 & 0.18                         & 0.38                         & 0.56 \\
USDNOK    & \cellcolor[HTML]{34FF34}0.03 & 0.12                         & 0.28                         & 0.37                         & 0.62 \\
USDSEK    & \cellcolor[HTML]{34FF34}0.05 & 0.22                         & 0.33                         & 0.44                         & 0.67
\end{tabular}
\caption{p values of the Dickey Fuller test on the At The Money Volatility series, for different assets and different maturities.}
\label{pvalues}
\end{table}
As we can see in Table \ref{pvalues}, for maturities larger than $6$ months, the Dickey Fuller test "shows" that there is no stationarity of the process, and so the time series are not mean reverting. Since the most liquid Variance Swaps are the ones with $1$ year maturity, we understand that trading on single assets is not a good idea, so it would be interesting to study the possible behavior of combination of Variance Swaps.
\subsection{Spread on Variance Swaps}
We now want to test whether spreads of volatilities might exhibit better stationarity properties than for the single vol assets. We focus on $1$ year volatilities. \\
We will apply the Dickey Fuller test on the spreads of the $1$ year  vol: here for simplicity we will consider the case where the spread is
$$
\Delta^{1,2}_t = y_t^1 - y_t^2,
$$
i.e., with a relative $\beta$ of $1$. A more sophisticated choice could be to use a different $\beta$ that could permit to balance the weights of the two Variance Swap / ATM volatilities series.  This is very important in the cross asset scenario: in fact, different asset classes present different ATM volatility levels, and then different Variance Swap levels. Equity volatilities are higher than FX volatilities, and choosing a simple spread with $\beta = 1$ would keep the properties of the equity series prominent with respect to the FX series. Nevertheless, in order to simplify the computation and the methodologies, we will keep $\beta=1$.\\
\begin{table}[H]
\centering
\begin{tabular}{c | c c c c c c c c c}
          & SPX                          & Eurostoxx                    & Footsie                      & Nikkei                       & Swiss                        & Canada                       & Australia                    & Norway                       & Sweden                       \\
          \hline
SPX       & NA                           & \cellcolor[HTML]{34FF34}0.01 & \cellcolor[HTML]{34FF34}0.01 & \cellcolor[HTML]{34FF34}0.01 & 0.12                         & \cellcolor[HTML]{34FF34}0.01 & \cellcolor[HTML]{34FF34}0.01 & 0.32                         & \cellcolor[HTML]{34FF34}0.01 \\
Eurostoxx & \cellcolor[HTML]{34FF34}0.01 & NA                           & \cellcolor[HTML]{34FF34}0.02 & \cellcolor[HTML]{34FF34}0.01 & 0.08                         & \cellcolor[HTML]{34FF34}0.01 & \cellcolor[HTML]{34FF34}0.01 & 0.31                         & \cellcolor[HTML]{34FF34}0.01 \\
Footsie   & \cellcolor[HTML]{34FF34}0.01 & \cellcolor[HTML]{34FF34}0.02 & NA                           & \cellcolor[HTML]{34FF34}0.01 & \cellcolor[HTML]{34FF34}0.01 & \cellcolor[HTML]{34FF34}0.03 & \cellcolor[HTML]{34FF34}0.01 & \cellcolor[HTML]{34FF34}0.04 & \cellcolor[HTML]{34FF34}0.01 \\
Nikkei    & \cellcolor[HTML]{34FF34}0.01 & \cellcolor[HTML]{34FF34}0.01 & \cellcolor[HTML]{34FF34}0.01 & NA                           & \cellcolor[HTML]{34FF34}0.01 & \cellcolor[HTML]{34FF34}0.01 & \cellcolor[HTML]{34FF34}0.01 & \cellcolor[HTML]{34FF34}0.01 & \cellcolor[HTML]{34FF34}0.01 \\
Swiss     & \cellcolor[HTML]{FFFFFF}0.12 & \cellcolor[HTML]{FFFFFF}0.08 & \cellcolor[HTML]{34FF34}0.01 & \cellcolor[HTML]{34FF34}0.01 & NA                           & \cellcolor[HTML]{34FF34}0.03 & \cellcolor[HTML]{34FF34}0.01 & 0.27                         & \cellcolor[HTML]{34FF34}0.01 \\
Canada    & \cellcolor[HTML]{34FF34}0.01 & \cellcolor[HTML]{34FF34}0.01 & \cellcolor[HTML]{34FF34}0.03 & \cellcolor[HTML]{34FF34}0.01 & \cellcolor[HTML]{34FF34}0.03 & NA                           & \cellcolor[HTML]{34FF34}0.01 & 0.08                         & \cellcolor[HTML]{34FF34}0.01 \\
Australia & \cellcolor[HTML]{34FF34}0.01 & \cellcolor[HTML]{34FF34}0.01 & \cellcolor[HTML]{34FF34}0.01 & \cellcolor[HTML]{34FF34}0.01 & \cellcolor[HTML]{34FF34}0.01 & \cellcolor[HTML]{34FF34}0.01 & NA                           & 0.19                         & \cellcolor[HTML]{34FF34}0.01 \\
Norway    & 0.32                         & 0.31                         & \cellcolor[HTML]{34FF34}0.04 & \cellcolor[HTML]{34FF34}0.01 & 0.27                         & 0.08                         & 0.19                         & NA                           & \cellcolor[HTML]{34FF34}0.01 \\
Sweden    & \cellcolor[HTML]{34FF34}0.01 & \cellcolor[HTML]{34FF34}0.01 & \cellcolor[HTML]{34FF34}0.01 & \cellcolor[HTML]{34FF34}0.01 & \cellcolor[HTML]{34FF34}0.01 & \cellcolor[HTML]{34FF34}0.01 & \cellcolor[HTML]{34FF34}0.01 & \cellcolor[HTML]{34FF34}0.01 & NA                          
\end{tabular}
\caption{p values of the Dickey Fuller test on spreads of $1$ year At The Money Volatility, Equity space.}
\label{dfequity}
\end{table}

\begin{table}[H]
\footnotesize
\centering
\begin{tabular}{c | c c c c c c c c c}
      & EURUSD                       & GBPUSD                       & USDJPY                       & USDCHF                       & USDCAD                       & AUDUSD                       & NZDUSD                       & USDNOK                       & USDSEK                       \\
       \hline
EURUSD & NA                           & 0.21                         & \cellcolor[HTML]{34FF34}0.01 & \cellcolor[HTML]{34FF34}0.02 & 0.34                         & \cellcolor[HTML]{34FF34}0.01 & 0.06                         & 0.19                         & 0.49                         \\
GBPUSD & 0.21                         & NA                           & \cellcolor[HTML]{34FF34}0.01 & 0.19                         & \cellcolor[HTML]{34FF34}0.02 & \cellcolor[HTML]{34FF34}0.01 & \cellcolor[HTML]{34FF34}0.01 & \cellcolor[HTML]{34FF34}0.01 & 0.30                         \\
USDJPY & \cellcolor[HTML]{34FF34}0.01 & \cellcolor[HTML]{34FF34}0.01 & NA                           & \cellcolor[HTML]{34FF34}0.01 & \cellcolor[HTML]{34FF34}0.03 & \cellcolor[HTML]{34FF34}0.05 & \cellcolor[HTML]{34FF34}0.04 & \cellcolor[HTML]{34FF34}0.01 & \cellcolor[HTML]{34FF34}0.03 \\
USDCHF & \cellcolor[HTML]{34FF34}0.02 & 0.19                         & 0.01                         & NA                           & 0.36                         & \cellcolor[HTML]{34FF34}0.04 & 0.07                         & 0.16                         & 0.32                         \\
USDCAD & 0.34                         & \cellcolor[HTML]{34FF34}0.02 & \cellcolor[HTML]{34FF34}0.03 & 0.36                         & NA                           & \cellcolor[HTML]{34FF34}0.02 & \cellcolor[HTML]{34FF34}0.02 & \cellcolor[HTML]{34FF34}0.03 & 0.12                         \\
AUDUSD & \cellcolor[HTML]{34FF34}0.01 & \cellcolor[HTML]{34FF34}0.01 & \cellcolor[HTML]{34FF34}0.05 & \cellcolor[HTML]{34FF34}0.04 & \cellcolor[HTML]{34FF34}0.02 & NA                           & \cellcolor[HTML]{34FF34}0.01 & \cellcolor[HTML]{34FF34}0.01 & \cellcolor[HTML]{34FF34}0.01 \\
NZDUSD & 0.06                         & \cellcolor[HTML]{34FF34}0.01 & \cellcolor[HTML]{34FF34}0.04 & 0.07                         & \cellcolor[HTML]{34FF34}0.02 & \cellcolor[HTML]{34FF34}0.01 & NA                           & \cellcolor[HTML]{34FF34}0.01 & \cellcolor[HTML]{34FF34}0.01 \\
USDNOK & 0.19                         & \cellcolor[HTML]{34FF34}0.01 & \cellcolor[HTML]{34FF34}0.01 & 0.16                         & \cellcolor[HTML]{34FF34}0.03 & \cellcolor[HTML]{34FF34}0.01 & \cellcolor[HTML]{34FF34}0.01 & NA                           & 0.08                         \\
USDSEK & 0.49                         & 0.30                         & \cellcolor[HTML]{34FF34}0.03 & 0.32                         & 0.12                         & \cellcolor[HTML]{34FF34}0.01 & \cellcolor[HTML]{34FF34}0.01 & 0.08                         & NA                          
\end{tabular}
\caption{p values of the Dickey Fuller test on spreads of $1$ year At The Money Volatility, FX space.}
\label{dffx}
\end{table}

\begin{table}[H]
\footnotesize
\centering
\begin{tabular}{c | c c c c c c c c c}
          & EURUSD                       & GBPUSD                       & USDJPY                       & USDCHF                       & USDCAD                       & AUDUSD                       & NZDUSD                       & USDNOK                       & USDSEK                       \\
          \hline
SPX       & 0.17                         & 0.16                         & 0.20                         & 0.25                         & 0.10                         & 0.05                         & 0.10                         & 0.10                         & 0.09                         \\
Eurostoxx & \cellcolor[HTML]{34FF34}0.01 & \cellcolor[HTML]{34FF34}0.01 & \cellcolor[HTML]{34FF34}0.04 & \cellcolor[HTML]{34FF34}0.02 & \cellcolor[HTML]{34FF34}0.01 & \cellcolor[HTML]{34FF34}0.01 & \cellcolor[HTML]{34FF34}0.01 & \cellcolor[HTML]{34FF34}0.01 & \cellcolor[HTML]{34FF34}0.01 \\
Footsie   & \cellcolor[HTML]{34FF34}0.02 & \cellcolor[HTML]{34FF34}0.01 & 0.10                         & \cellcolor[HTML]{34FF34}0.04 & \cellcolor[HTML]{34FF34}0.01 & \cellcolor[HTML]{34FF34}0.02 & \cellcolor[HTML]{34FF34}0.02 & \cellcolor[HTML]{34FF34}0.02 & \cellcolor[HTML]{34FF34}0.04 \\
Nikkei    & \cellcolor[HTML]{34FF34}0.01 & \cellcolor[HTML]{34FF34}0.01 & \cellcolor[HTML]{34FF34}0.02 & \cellcolor[HTML]{34FF34}0.02 & \cellcolor[HTML]{34FF34}0.01 & \cellcolor[HTML]{34FF34}0.01 & \cellcolor[HTML]{34FF34}0.01 & \cellcolor[HTML]{34FF34}0.01 & \cellcolor[HTML]{34FF34}0.01 \\
Swiss     & 0.09                         & \cellcolor[HTML]{34FF34}0.04 & 0.19                         & 0.15                         & \cellcolor[HTML]{34FF34}0.02 & \cellcolor[HTML]{34FF34}0.03 & \cellcolor[HTML]{34FF34}0.03 & 0.05                         & 0.09                         \\
Canada    & 0.17                         & 0.14                         & 0.26                         & 0.24                         & 0.13                         & 0.09                         & 0.11                         & 0.12                         & 0.12                         \\
Australia & 0.17                         & 0.09                         & 0.24                         & 0.21                         & 0.06                         & \cellcolor[HTML]{34FF34}0.04 & 0.07                         & 0.09                         & 0.13                         \\
Norway    & 0.23                         & 0.17                         & 0.26                         & 0.25                         & 0.27                         & 0.27                         & 0.17                         & 0.25                         & 0.27                         \\
Sweden    & \cellcolor[HTML]{34FF34}0.01 & \cellcolor[HTML]{34FF34}0.01 & \cellcolor[HTML]{34FF34}0.02 & \cellcolor[HTML]{34FF34}0.01 & \cellcolor[HTML]{34FF34}0.01 & \cellcolor[HTML]{34FF34}0.01 & \cellcolor[HTML]{34FF34}0.01 & \cellcolor[HTML]{34FF34}0.01 & \cellcolor[HTML]{34FF34}0.01
\end{tabular}
\caption{p values of the Dickey Fuller test on spreads of $1$ year At The Money Volatility, Cross - Asset space.}
\label{dfxasset}
\end{table}
It is easy to check that spreads exhibit much stronger stationarity properties, in that most of
the spreads now pass the stationarity test. The results above suggest that taking a position on spreads of cross-asset volatilities should
permit magnifying the mean reversion strategies already evident for the volatility variables.
\subsubsection{A case study}
Motivated by Tables \ref{dfequity}, \ref{dffx} and \ref{dfxasset}, we study the spread between SPX and Eurostoxx and between SPX and Footsie. This two spreads show mean reversion from the Dickey Fuller test.\\
We compute the half life of the two spreads, and we compare them with the half life of the single assets:
\begin{table}[H]
\centering
\begin{tabular}{c|c|c|c|c}
SPX    & Eurostoxx & Footsie & SPX - Eurostoxx & SPX - Footsie \\
\hline
138.89 & 26.84     & 38.15   & 2.61            & 2.16         
\end{tabular}
\caption{Half lives in days of $12$ months At The Money Volatilities and their spreads}
\label{halflives}
\end{table}
\begin{figure}[H]
  \includegraphics[width=\linewidth]{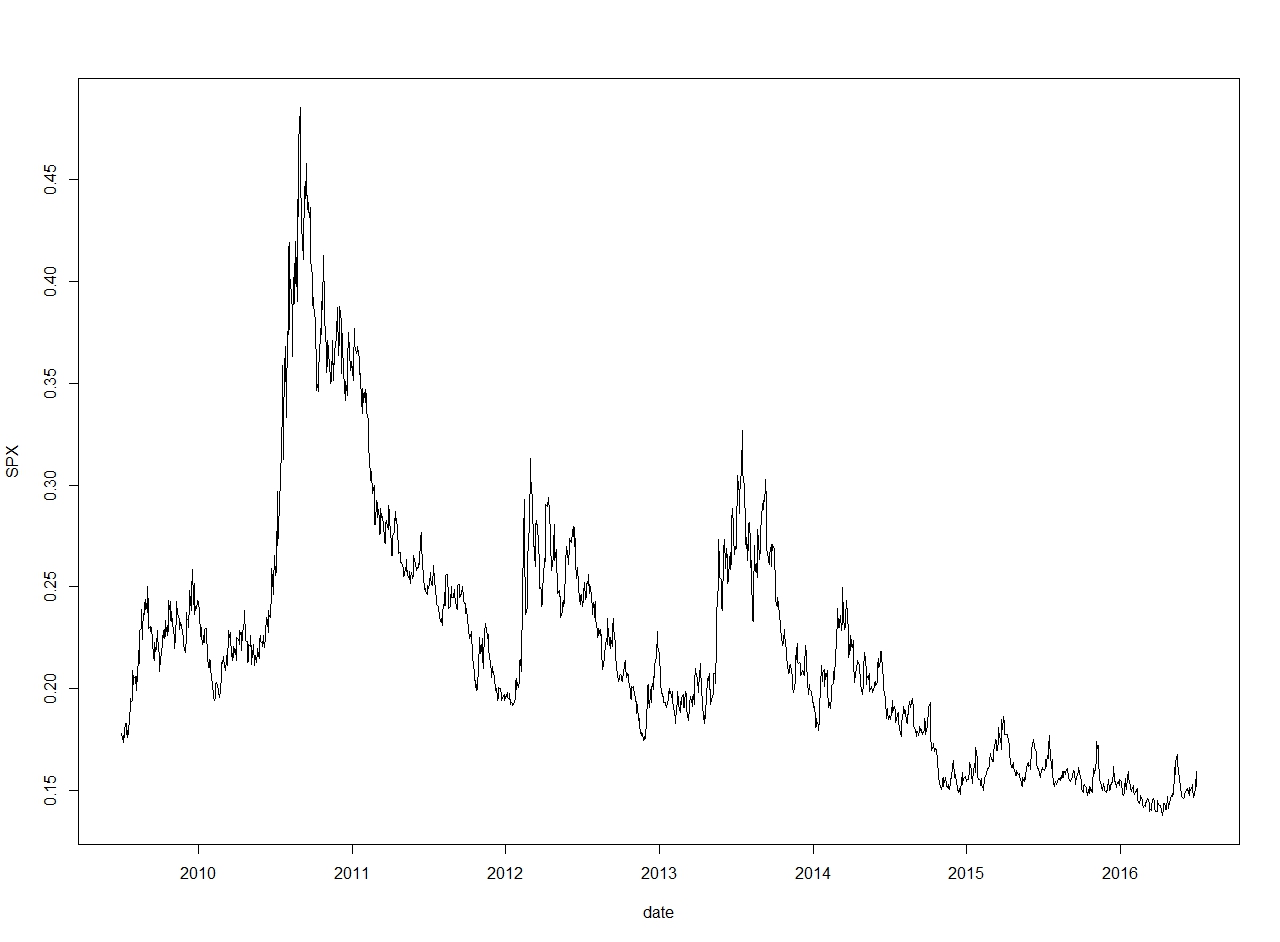}
  \caption{Time series of the $12$ months At The Money Volatility of the SPX index, from July 1, 2009 to July, 1 2016.}
  \label{SPX}
\end{figure}
The time series of the $12$ ATM Volatility does not present mean reversion properties, as can we see in Figure \ref{SPX} and in Tables \ref{pvalues} and \ref{halflives}. Moreover, it presents peaks in correspondence of periods of stress, as 2011 and 2013. The use of spreads, in addition of the already shown property of magnification of the mean return properties, will remove the peaks of the time series.
\begin{figure}[H]
  \includegraphics[width=\linewidth]{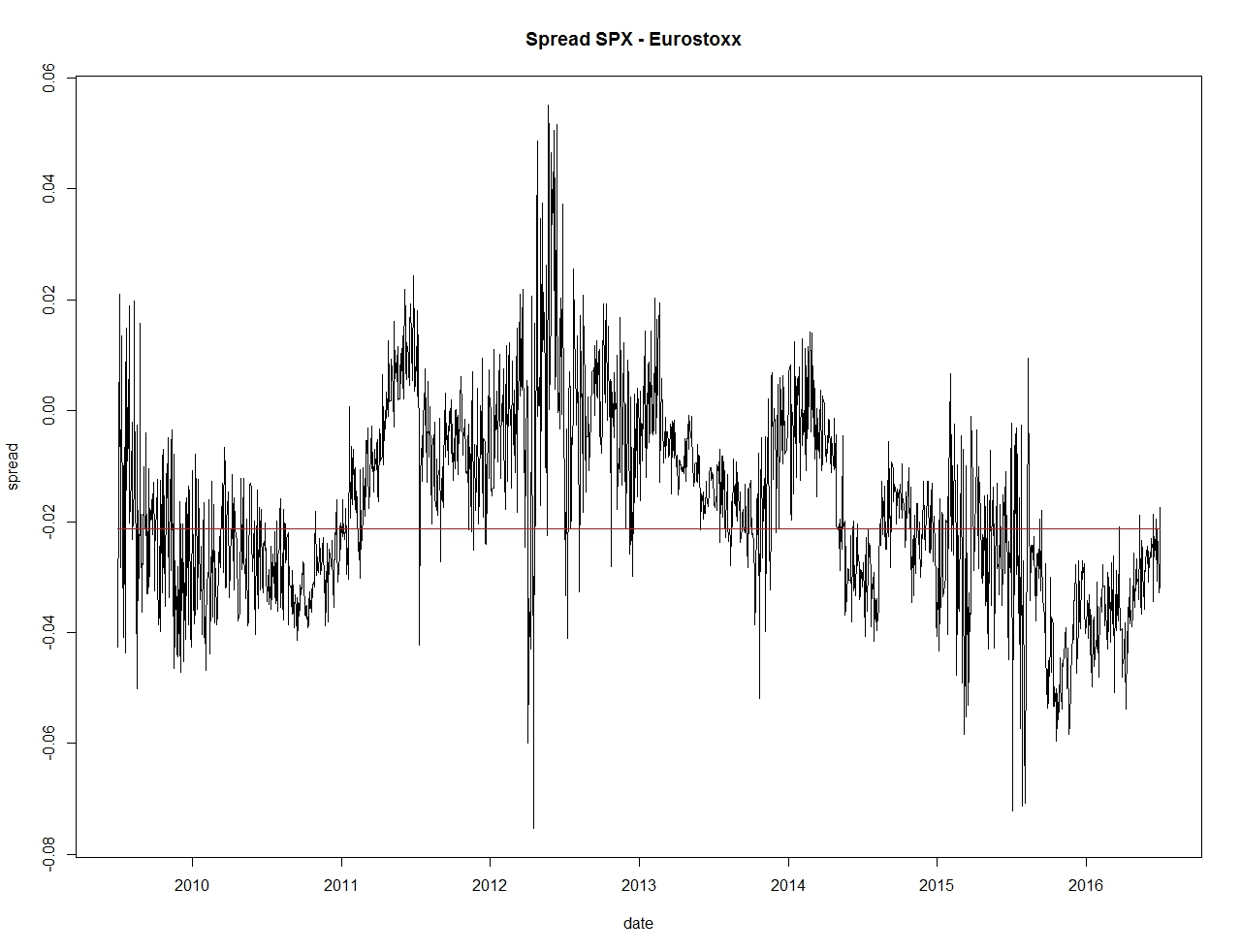}
  \caption{Time series of the $12$ months At The Money Volatility spreads between the SPX and the Eurostoxx indexes, from July 1, 2009 to July, 1 2016.}
  \label{spread_spx_euro}
\end{figure}
\begin{figure}[H]
  \includegraphics[width=\linewidth]{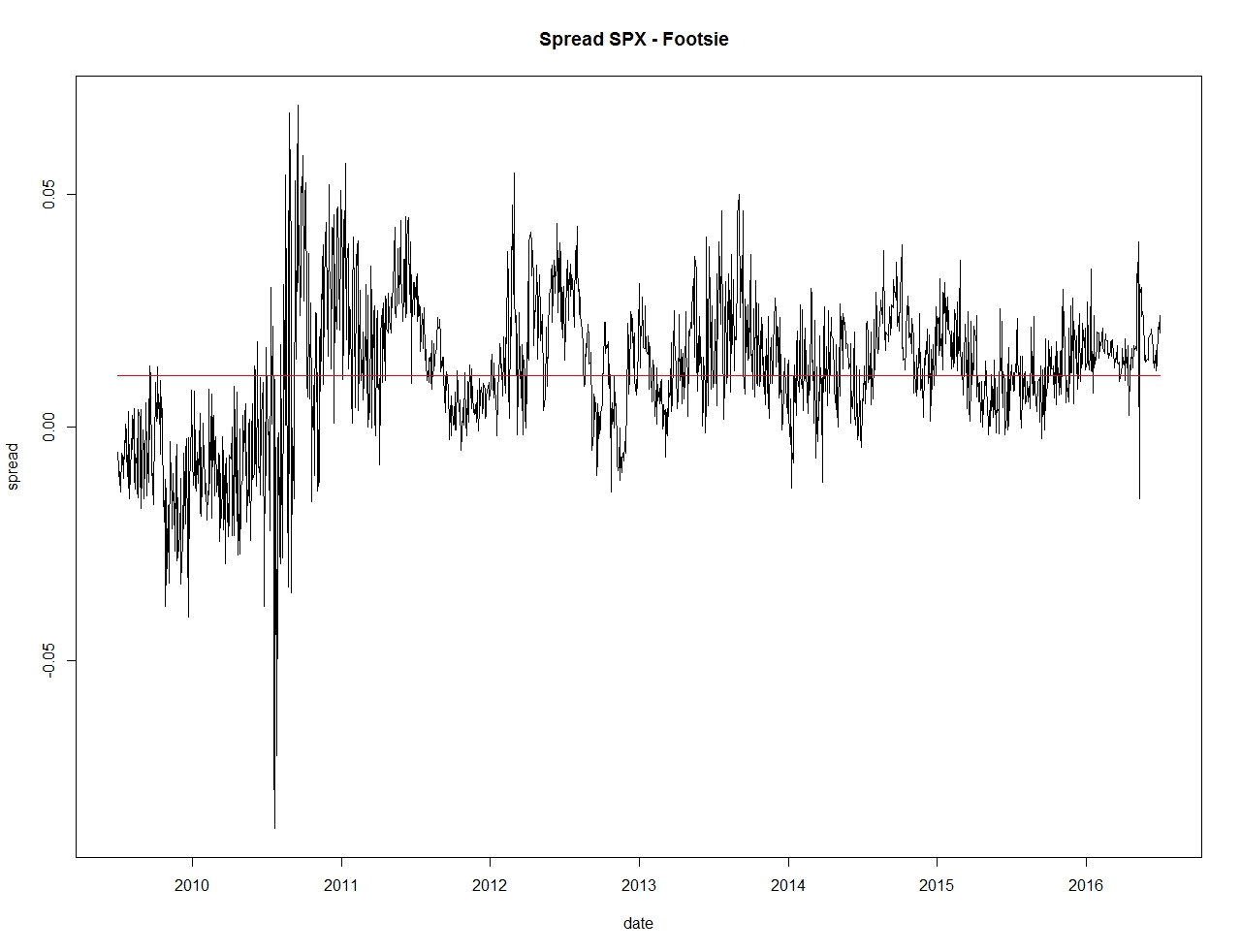}
  \caption{Time series of the $12$ months At The Money Volatility spreads between the SPX and the Footsie indexes, from July 1, 2009 to July, 1 2016.}
  \label{spread_spx_footsie}
\end{figure}
Even if the mean values are low, Figures \ref{spread_spx_euro} and \ref{spread_spx_footsie} show that it is possible to remove the spikes in correspondence of high volatility periods, and it is easily seen that the mean reversion properties are magnified, since we can see the almost periodic behavior of the spreads time series. This observation give now a good hint on how to possible trade this couples of Variance Swaps, in order to exploit all the mean reversion properties. This would not have been possible if we had focus on only one asset, since the mean reverting trading strategies would have not worked.
\subsection{Final analysis}
As we have seen in the previous sections, there are two possible ways to implement investment strategies based on variance swap: study the difference between the implied and statistic parameters, and exploit the mean reverting characteristics of the spreads.
\subsubsection{Risk premium on Variance Swaps}
The market premium on a vanilla option, say a Call option, is given by the difference of the market price and the price given by the Black Scholes formula where the volatility is given by a statistical estimator. Since the Black Scholes price is a bijective function of the volatility, the market premium is directly given by the difference between the implied volatility and the statistical volatility. \\ 
This relation is easy to understand and to implement in the case of plain vanilla options since, even if the Black Scholes model is not a good model, it is a common assumption that implied volatility in the Black Scholes formulas is "the wrong number to put in the wrong formula to obtain the right price". Things becomes more complicated with more exotic derivatives, in particular with Variance Swap, as in our case. We can assume that the SABR formula \eqref{KvarSABR} is the equivalent of the Black Scholes formula. In this case, the "wrong" numbers to put in the formula would be the calibrated values of the observed variance $\alpha$ and the volvol parameter $\nu$.\\ 
$\alpha$, if not calibrated, can be easily estimated by the ATM implied volatility of $1$ week plain vanilla options, while the volvol parameter can be obtained inverting formula \eqref{KvarSABR}, if the series of the Variance Swap prices is available.\\ 
This has to be compared with the statistical variance and the statistical volvol. This elements are usually obtained using an EWMA with $\lambda=0.94$ from, respectively, the stock price series and the (estimated) volatility series.
The market premium of a variance swap becomes, under this assumptions, the sum of the premium given by the volatility and the premium given by the volvol parameter. The advantage of the strategies using Variance Swap is that, even if we add more complexity to the pricing and hedging procedures, we add a component in the market premium, so even when the volatility premium is low (so we do not have signals on whether to buy or sell plain vanilla options), there is still the possibility to exploit the volvol premium.
\subsubsection{Mean reversion on spreads of Variance Swaps}
Mean reversion properties of Variance Swaps have been studied in the previous section: a mean reverting process can give trading signals on when to buy and when to sell an asset or a derivative. When the price is high and we know that, according to the model, the price is lowering, then it is a good idea to sell, because it would not be possible to extract more value from the carry. Similarly, when the price is low and the models suggests that the series is reverting to the mean, then it is a good choice to buy, since the wait would only make the price of the asset / derivative higher.\\ 
In case of derivatives, it use useful to study the half life of the process. In the case of derivatives, if the half life is comparable to the maturity of the option, than this strategy of buying / selling would not be applicable, because the time taken by the process to return to his mean would overcome the maturity date.\\ 
Motivated by this theoretical and empirical observation, the spread of Variance Swaps is a great way to highlight mean reverting characteristic and to implement trading strategies. In fact, as already seen, the series of the spread is much more mean reverting, with the half life usually under five trading days. Even if we did not consider the presence of trading costs, the possibility of obtain value from the strategy on a weekly basis is interesting and suggests full attention.\\ 
In this work we did not implement any trading strategy. This is due to the fact that the scope of the Research team is to suggest new technologies and new trading ideas to the investors and to the trading desks. The choice of the derivatives to trade, the asset class, the most interesting strategy, is not the job of the research department, so there will not be any market implementation in this report.

\newpage
\phantom{bla}

\end{document}